\documentclass[american,nofootinbib]{revtex4-2}
\usepackage[T1]{fontenc}
\usepackage[utf8]{inputenc}
\usepackage{color}
\usepackage{babel}
\usepackage{array}
\usepackage{booktabs}
\usepackage{amsmath}
\usepackage{amssymb}
\usepackage{graphicx}
\usepackage[pdfusetitle,
 bookmarks=true,bookmarksnumbered=false,bookmarksopen=false,
 breaklinks=false,pdfborder={0 0 1},backref=false,colorlinks=true]
 {hyperref}
\hypersetup{
 urlcolor=blue, citecolor=blue, linkcolor=blue}

\makeatletter

\providecommand{\tabularnewline}{\\}

\makeatother

\begin{document}
\title{Flux-driven turbulent transport using penalisation in the Hasegawa-Wakatani
system}
\author{Pierre L. Guillon$^{1,2}$, Özgür D. Gürcan$^{1}$, Guilhem Dif-Pradalier$^{3}$,
Yanick Sarazin$^{3}$ and Nicolas Fedorczak$^{3}$}
\affiliation{$^{1}$Laboratoire de Physique des Plasmas, CNRS, Ecole Polytechnique,
Sorbonne Université, Université Paris-Saclay, Observatoire de Paris,
F-91120 Palaiseau, France}
\affiliation{$^{2}$ENPC, Institut Polytechnique de Paris, Marne-la-Vallée, France}
\affiliation{$^{3}$CEA, IRFM, F-13108 Saint-Paul-lez-Durance, France}
\begin{abstract}
First numerical results from the newly-developed pseudo-spectral code
P-FLARE (Penalised FLux-driven Algorithm for REduced models) are presented.
This flux-driven turbulence/transport code uses a pseudo-spectral
formulation with the penalisation method in order to impose radial
boundary conditions. Its concise, flexible structure allows implementing
various quasi-two dimensional reduced fluid models in flux-driven formulation.
Here, results from simulations of the modified Hasegawa-Wakatani system
are discussed, where particle transport and zonal flow formation,
together with profile relaxation, are studied. It is shown that coupled
spreading/profile relaxation that one obtains for this system is consistent
with a simple one dimensional model of coupled spreading/transport
equations. Then, the effect of a particle source is investigated,
which results in the observation of sandpile-like critical behaviour.
The model displays profile stiffness for certain parameters, with
very different input fluxes resulting in very similar mean density
gradients. This is due to different zonal flow levels around the critical
value for the control parameter (\emph{i.e.} the ratio of the adiabaticity
parameter to the mean gradient) and the existence for this system
of a hysteresis loop for the transition from 2D turbulence to a zonal
flow dominated state.
\end{abstract}
\maketitle

\section{Introduction}

In magnetised plasmas, large temperature and density gradients drive
turbulent transport, which affects the confinement capability of magnetic
fusion devices. The two paradigms that aim to address the effect of
turbulence on heat and particle transport are \emph{fixed gradient}
or \emph{flux-driven} approaches. In the former, a mean profile is
imposed on the system, which provides a mean temperature or density
gradient, constant in time. This gradient then triggers a linear instability
that saturates by generating turbulence, which in turn creates turbulent
heat or particle fluxes. But in this approach, the imposed profile
is not affected by the turbulent fluxes through any kind of transport
equation. The fixed gradient framework relies on the separation of
time and spatial scales, with the hypothesis that turbulence develops
much faster than typical response times of the profile, which is assumed
to be ``frozen''. Fixed gradient systems are useful to study the formation
and the saturation mechanisms of turbulence, and can help estimate
the turbulent flux (or turbulence level) at a given mean density gradient.
However, the interplay between turbulence and transport is completely
lost, even though one may have zonal flows and zonal corrugations
that appear as perturbations on top of the fixed gradient.

In contrast, flux driven systems do not assume this scale separation
\citep{garbet:1998,gillot:2023,panico:2025}. They feature a transport
equation on the profile, which allows the mean gradients to self-consistently
evolve in response to the turbulent fluxes, along with additional
sources and sinks. As a result, they offer a more realistic framework,
where it is the particle and heat sources and not the gradients that
are externally imposed on the system. However, from a computational
perspective, flux-driven systems are usually more complicated to implement
and numerical simulations are more costly, partly because of the need
for reaching flux equilibrium.

Ideally, tokamak turbulence should be described using gyrokinetics
to incorporate all kinetic effects, such as Landau damping, instabilities
involving energetic and trapped particles, electromagnetic effects
including large scale Alfvèn modes, as well as small scale electromagnetic
turbulence such as micro-tearing. But, since such gyrokinetic codes
require high computational time and resources, and they come with
a complex theoretical framework, the use of \emph{reduced fluid models}
may often be prefered, especially if the goal is to understand complex
nonlinear feedback mechanisms such as the interplay between turbulence,
transport and zonal flows. These models are obtained by taking the
first few moments of the (gyro-) kinetic equations, and display already
quite rich behaviours including zonal flows or avalanches. Considering
two-dimensions, the minimal model which features turbulence generation
from a linear instability driven by a mean-gradient, and its self-organisation
into zonal flows, is the Hasegawa-Wakatani model \citep{hasegawa:1983},
which has been widely studied both theoretically and numerically for
the past 40 years \citep{hu:1997,camargo:1998,numata:2007,anderson:2017,kim:2019,heinonen:2020,gurcan:2022}.

Reduced fluid models generally feature a small number of equations,
and here we consider quasi-2D models, since the wavenumber of the
fluctuations parallel to the magnetic field lines $k_{\parallel}$
is much lower than in the perpendicular direction $k_{\perp}$. In
this approximation, these models are similar to the 2D incompressible
Navier-Stokes system. Performing fixed gradient simulations of these
systems in a 2D, periodic, slab geometry can be reasonably and efficiently
achieved using a pseudo-spectral formulation relying on \emph{fast
Fourier transforms}. This makes it natural to have \emph{periodic
boundary conditions}, which means that the mean gradient must be extracted
from the profile and used as a constant parameter in such a system,
while the remaining part constitutes turbulent fluctuations, which
are taken to be periodic in both directions. In contrast, flux-driven
systems do not normally use the pseudo-spectral method as they do
not have periodic boundary conditions in the radial direction. They
are instead often solved using finite difference numerical schemes,
which can be slower to compute and less precise, especially for linear
terms involving higher order derivatives.

To further improve computation time, one may impose further reductions,
\emph{e.g.} as in the Tokam1D model \citep{panico:2025}, where radial/zonal
profiles are completely resolved, but only one poloidal mode is retained.
Although such models can describe complex phenomena such as zonal
flow generation and avalanche formation, they lack the fluctuating
turbulent-turbulent nonlinear interaction term, which can be recovered,
at least partially, either by using eddy viscosity closure terms \citep{carnevale:1982,krommes:2002}
or by adding more poloidal modes and their interactions \citep{terry:1983,guillon:2025}.
However, adding even a few poloidal modes considerably complicates
the basic structure of such a reduced model, while in a pseudo-spectral
formulation one can decide how many poloidal modes to incorporate
freely, without any difficulty.

To benchmark these further reduced models, one needs to compare their
results to actual direct numerical simulations (DNS) of flux-driven
plasma fluid models. For this purpose, and to study flux-driven nonlocal
transport in reduced fluid models in general, we developed the P-FLARE
code \citep{guillon:2025a}, which stands for Penalised FLux-driven
Algorithm for REduced models. The code uses the penalisation method
introduced by Ref. \citealp{angot:1999}, which works
by constructing \emph{buffer zones} at both ends of the radial domain.
In these buffer regions, turbulent fluctuations are suppressed, and
the radial profiles are modified so that they can be written as a
combination of a periodic part and a jump, which allows us to compute
their fast Fourier transforms and use the pseudo-spectral method.
Thus, starting from a pseudo-spectral implementation of a fixed gradient
system, we can easily obtain the implementation for the corresponding
flux-driven system, while keeping the advantages of using fast Fourier
transforms in order to perform high (\emph{i.e. }$4096\times4096$
padded) resolution simulations that run at a comparable speed to its
fixed-gradient pseudo-spectral version.

As a first application of the code, we perform numerical simulations
of the flux-driven Hasegawa-Wakatani (FDHW) system. However, the flexible
formulation of the code allows us to easily apply it to other similar
reduced fluid models, such as 2D models for edge turbulence \citep{sarazin:1998,sarazin:2021,panico:2025}
which can include interchange, finite Larmor radius effects and the
effect of the diamagnetic stress, or fluid ITG/ETG models \citep{horton:1981,horton:1988,ivanov:2020}
for core turbulence.

In order to demonstrate the capabilities of the code, we first consider
a density profile which is steep in the inner radius and flat in the
outer radius, as the initial condition, and study its relaxation in
the absence of source. At short times, we observe \emph{``pair formation''}
around the steepest part of the initial profile, with density bumps
and holes growing linearly and then the bump moving radially outward
while the hole moves inward. On longer time scales, the profile gradually
relaxes, while the turbulent region, initially excited in the vicinity
of the steep part of the density profile, spreads towards the flat
region. In order to reproduce this relaxation, coupled to turbulent
spreading, we use a 1D model, which consists of an equation for the
turbulent kinetic energy, coupled to a transport equation for the
density profile. Both equations feature a nonlinear diffusive term,
in which the nonlinear diffusion coefficient is proportional to the
turbulent kinetic energy. Such models have often been used in the
past, for studying profile relaxation and turbulence spreading together
or separately \citep{hahm:2004,gurcan:2005,naulin:2005,garbet:2007,miki:2012}.
In the case of the flux-driven Hasegawa-Wakatani system, we observe
that the reduced model reproduces both qualitatively and quantitatively
the turbulent front propagation, which, for this particular problem,
is found to be slightly subdiffusive.

At the end of the simulation, the system is observed to be in a state
dominated by zonal flows. This is a manifestation of the transition
from 2D turbulence to zonal flows observed in the fixed gradient system
when the ratio between the adiabaticity parameter $C$ and the mean
gradient $\kappa$ is varied \citep{numata:2007,grander:2024,guillon:2025}.
The critical point for this transition, which is around $C/\kappa=0.1$,
is reached at some point as the density profile relaxes and the mean
density gradient $\kappa$ decreases in time. Once the system transitions
to a state dominated by zonal flows, the turbulent flux is suppressed
and a \emph{marginally} stable state is established, which does not result in a completely flattened final state.

In order to compare both runtimes and physical results with the more conventional
finite difference formulation, we have benchmarked our relaxation
simulation against the BOUT++ code \citep{dudson:2009}, and the results,
which can be found in Appendix \ref{sec:Benchmark}, show reasonable
agreement.

Lastly, as an actual example of flux-driven simulations, we explore
the effect of a localised particle source on the final state. Starting
from the same initial saturated state, we find very different final
states, by varying the source amplitude. Due to the existence of a
marginally stable state dominated by zonal flows, the density profile
exhibits behaviour somewhat reminiscent of sandpile models in self-organised
criticality \citep{bak:1987,hwa:1992}. To be more precise, the existence
of a hysteresis loop suggestive of a first order transition, and the
fact that the critical state is not the threshold of a linear instability
but of a nonlinear transition, suggests that we have what is called
self-organised bistability \citep{gil:1996,disanto:2016}. Another
consequence of the presence of this hysteresis loop is that it causes
profile stiffness \citep{wolf:2003,garbet:2004,mantica:2009}, in
the sense that within some range of parameters close to the transition
value, the simulations with very different imposed fluxes result in
very similar final profiles. We argue that this is mainly due to differences
in zonal flow levels between these simulations.

The rest of the article is organised as follows. In section \ref{sec:FDHW},
we define the flux-driven Hasegawa-Wakatani model starting from the
original fixed gradient model. Then, in section \ref{sec:P-FLARE},
we detail the penalisation method that we use in the P-FLARE code,
which requires buffer regions, modifying the radial profiles
so that they become periodic in these regions, and prescribing boundary
conditions. The relaxation of the density profile is studied in section
\ref{sec:spread_transport}, along with turbulence spreading, which
is reproduced by the 1D reduced model. Finally, in section \ref{sec:sources},
we discuss the impact of the localised particle source, which displays
sandpile-like behaviour and profile stiffness.

\section{The flux-driven Hasegawa-Wakatani model}\label{sec:FDHW}

In this section, we present the derivation of the particle flux-driven
Hasegawa-Wakatani system, which we will call FDHW from now on. First,
we recall the classical, fixed gradient equations.

\subsection{The fixed gradient Hasegawa-Wakatani model}

We consider the Hasegawa-Wakatani model in a 2D slab $\left(\mathbf{e}_{x},\mathbf{e}_{y}\right)$
of size $L_{x}\times L_{y}$ orthogonal to a uniform and constant
magnetic field $\mathbf{B}=B\mathbf{e}_{z}$, where $x$ and $y$
are respectively the radial and poloidal directions normalised to
the sound Larmor radius $\rho_{s}$. We first assume that the fluctuations
are periodic in both axes.

The modified Hasegawa-Wakatani equations are \citep{hasegawa:1983,numata:2007,guillon:2025}

\begin{subequations}
\begin{align}
\partial_{t}\Omega+\left[\phi,\Omega\right] & =C(\widetilde{\phi}-\widetilde{n})\,,\label{eq:hw-om}\\
\partial_{t}n+\left[\phi,n\right]+\kappa\partial_{y}\phi & =C(\widetilde{\phi}-\widetilde{n})\,,\label{eq:hw-n}
\end{align}
\end{subequations}
where $\phi$ is the electrostatic potential, $n$ the density perturbation,
and $\Omega=\nabla_{\perp}^{2}\phi$ corresponds to the vorticity.
The Poisson bracket $\left[\phi,A\right]=\partial_{x}\phi\partial_{y}A-\partial_{y}\phi\partial_{x}A$
represents the advection of $A$ by the $E\times B$ velocity $\mathbf{v}_{E\times B}=\mathbf{e}_{z}\times\nabla\phi$.
The adiabaticity parameter $C=\frac{k_{||}^{2}\rho_{s}^{2}B}{en_{0}\eta_{e,||}}$
, with $e$ the unit charge, $k_{||}$ the parallel wave-number of
fluctuations (along the magnetic field lines) and $\eta_{e,||}$ the
parallel resistivity, and the background density gradient $\kappa\equiv-\frac{1}{n_{r}}\frac{dn_{r}}{dx}$,
with $n_{r}$ being the background density profile normalised to a
constant density reference $n_{0}$, are constant parameters of the
model. Even though here, we consider the inviscid case for simplicity,
dissipation terms are generally introduced in both equations for numerical
studies. Note that we have also decomposed the perturbations $A$
into their zonal $\overline{A}$ (averaged over the poloidal direction)
and non zonal $\widetilde{A}$ components, with a poloidal average
defined as:
\begin{equation}
\overline{A}(x,t)=\langle A\rangle_{y}\equiv\frac{1}{L_{y}}\int_{0}^{L_{y}}A\,dy\,,\;\langle\widetilde{A}\rangle_{y}=0\,.\label{eq:zfturb}
\end{equation}

In fact, (\ref{eq:hw-n}) can also be written for the total density
$N(x,y,t)=n_{r}(x,t)+\widetilde{n}(x,y,t)$, normalised here to the
constant reference density $n_{0}$, so that:
\begin{equation}
\partial_{t}N+\left[\phi,N\right]=C(\tilde{\phi}-\tilde{n})\,,\label{eq:hw-ntot}
\end{equation}
where
\begin{equation}
n_{r}=n_{lin}+\overline{n}\label{eq:nrdef}
\end{equation}
 is the radial density profile, which combines the zonal density fluctuations
with the linear (\emph{i.e.} fixed gradient) background density profile
as:
\begin{equation}
n_{lin}(x,t)=-\kappa(x-L_{x})+n_{r}(x=L_{x})\,.\label{eq:nlindef}
\end{equation}
The Poisson bracket for $n_{lin}$ indeed yields $\left[\phi,n_{lin}\right]=-\partial_{x}n_{lin}\partial_{y}\phi=\kappa\partial_{y}\phi$.

In a fixed gradient formulation, since $\kappa$ is a constant, the
background density profile $n_{lin}$ is prescribed and does not change
in time. One may argue that this corresponds to imposing a complicated
form of source terms to the density profile, which compensates the
turbulent particle flux generated by the underlying instability \citep{garbet:1998},
and that the remaining zonal density is basically a radially periodic
perturbation, which cannot change the mean gradient.

\subsection{The flux-driven model}\label{subsec:eqs-FD}

In order to develop the flux-driven formulation, let us first consider
$n_{r}$ to be an arbitrary density profile subject to a transport
equation due to the turbulent particle flux.

\subsubsection{Decomposition of the radial density profile}

\begin{figure}
\centering{}\includegraphics[width=0.45\textwidth]{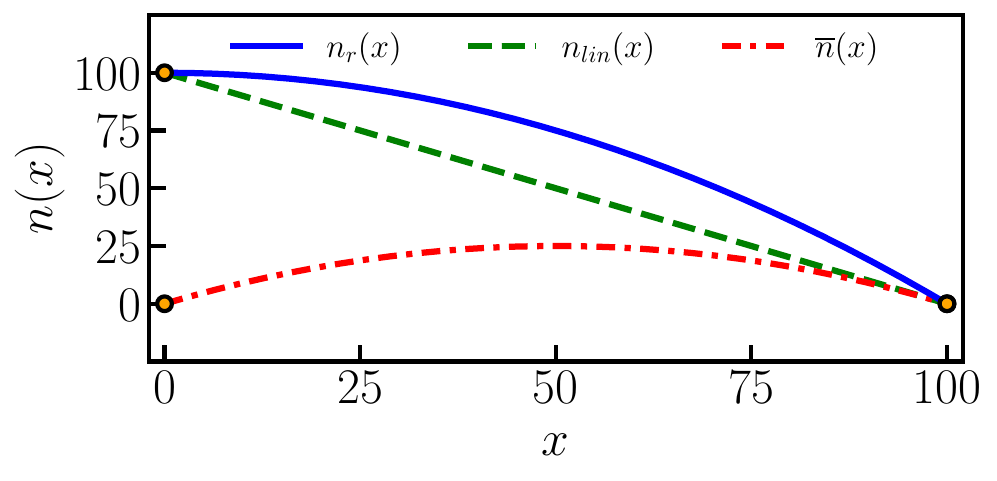}\caption{Decomposition of a given radial profile $n_{r}$ (blue) into its linear
component $n_{lin}$ (dashed green) and its zonal fluctuating component
$\overline{n}$ (dashdot red). Notice that using such a decomposition,
we find that the zonal profile is periodic at the edges, while its
derivative $\partial_{x}\overline{n}$ is not.}\label{fig:decomp}
\end{figure}

We can still decompose the density profile into a ``linear'' part
$n_{lin}$ and a radially periodic perturbation part $\overline{n}$,
which is just the zonal part of the density profile:
\begin{equation}
n_{r}(x,t)=n_{lin}(x,t)+\overline{n}(x,t)\,,\label{eq:nr}
\end{equation}
where
\begin{equation}
n_{lin}(x,t)\equiv-\kappa(t)(x-L_{x})+n_{r}(x=L_{x},t)\,,\label{eq:nlin}
\end{equation}
is no longer a constant in time, so that
\begin{equation}
\kappa(t)\equiv-\frac{n_{r}(x=L_{x},t)-n_{r}(x=0,t)}{L_{x}}\,,\label{eq:kap}
\end{equation}
is a function of time, defined for the radial profile at a given time,
and corresponds to the mean density gradient in the radial direction.
The decomposition is illustrated in figure \ref{fig:decomp}.

This decomposition is important because it enables us to obtain a
radially periodic perturbated part, \emph{i.e.} $\overline{n}$, from
which we will compute the fast Fourier transform, while the linear
part will correspond to the constant in space $\kappa(t)$ in Eq.
\ref{eq:hw-n}, which is now time dependent. Furthermore, this decomposition
may help us project what we know about the linear, zonal and transport
properties of the classic gradient-driven system into the flux-driven
version, where the linear parameter $\kappa$ will now evolve in time
in response to the turbulent particle transport. We know for example
from fixed gradient studies of the Hasegawa-Wakatani system \citep{numata:2007,grander:2024,guillon:2025}
that for $C/\kappa\lesssim0.1$, we get 2D like turbulence dominated
by eddies, while for $C/\kappa\gtrsim0.1$ zonal flows dominate. We
can argue that this observation should apply equally using a time
evolving local mean gradient in a flux driven system.

Notice that we can also define the zonal density to have a zero mean
value:
\begin{equation}
\overline{n}\equiv n_{r}-n_{lin}-\left\langle n_{r}-n_{lin}\right\rangle _{x}\,,\label{eq:zonal_0mean}
\end{equation}
where $\langle\cdot\rangle_{x}$ denotes averaging along the $x$
axis. Thus, we have
\begin{equation}
n_{r}=n_{lin}+\overline{n}+n_{m}\,,\label{eq:nrfddef}
\end{equation}
with $n_{m}\equiv\left\langle n_{r}-n_{lin}\right\rangle _{x}$, so
that $\overline{n}$ is really a zero mean, radially periodic perturbation
associated with the density profile, as it would be the case in a
fixed gradient model.

\subsubsection{The equations}

We can then write the time derivative of the total density $N(x,y,t)=n_{r}(x,t)+\widetilde{n}(x,y,t)$
using Eq. (\ref{eq:hw-ntot}) as: 
\begin{align}
\partial_{t}N+\left[\phi,N\right] & =\partial_{t}n_{r}+\partial_{t}\widetilde{n}+\left(\kappa(t)-\partial_{x}\overline{n}\right)\partial_{y}\phi+\left[\phi,\widetilde{n}\right]=C(\widetilde{\phi}-\widetilde{n})\,,\label{eq:ntotbis}
\end{align}
where we decomposed the Poisson bracket of the total density. Averaging
this along $y$, we get the evolution equation for the radial density
profile:
\begin{equation}
\partial_{t}n_{r}+\partial_{x}\Gamma_{n}=0\,\label{eq:hw-nr-raw}
\end{equation}
with $\Gamma_{n}$ the radial particle flux that is averaged along
$y$, defined as
\begin{equation}
\Gamma_{n}(x,t)\equiv\langle\widetilde{n}\widetilde{v}_{x}\rangle_{y}=-\langle\widetilde{n}\partial_{y}\widetilde{\phi}\rangle_{y}\,.\label{eq:GAM}
\end{equation}
Note that density source terms can be introduced in Eq. \ref{eq:hw-nr-raw}.
Subtracting the averaged equation (\ref{eq:hw-nr-raw}) from the complete
equation (\ref{eq:ntotbis}), we get the time evolution of the non-zonal
density fluctuations
\begin{equation}
\partial_{t}\widetilde{n}+\left(\kappa(t)-\partial_{x}\overline{n}\right)\partial_{y}\phi+[\phi,\widetilde{n}]-\langle[\widetilde{\phi},\widetilde{n}]\rangle_{y}=C(\widetilde{\phi}-\widetilde{n})\,,\label{eq:ntilde}
\end{equation}
so that we can write the complete flux-driven model as:

\begin{subequations}
\begin{align}
\partial_{t}\Omega+\left[\phi,\Omega\right] & =C(\widetilde{\phi}-\widetilde{n})+\nu\nabla^{2}\widetilde{\Omega}\,,\label{eq:hwfd-om}\\
\partial_{t}\widetilde{n}+\left(\kappa(t)-\partial_{x}\overline{n}\right)\partial_{y}\widetilde{\phi}+\partial_{x}\overline{\phi}\partial_{y}\widetilde{n}+[\widetilde{\phi},\widetilde{n}]-\langle[\widetilde{\phi},\widetilde{n}]\rangle_{y} & =C(\widetilde{\phi}-\widetilde{n})+D\nabla^{2}\widetilde{n}\,,\label{eq:hwfd-nturb}\\
\partial_{t}n_{r}+\partial_{x}\Gamma_{n} & =S_{n}(x,t)+D_{0}\partial_{x}^{2}n_{r}\,,\label{eq:hwfd-nr}
\end{align}
\end{subequations}
where $\kappa$ and $\Gamma_{n}$ are given by Eqs. \ref{eq:kap}
and \ref{eq:GAM} respectively. Note that viscosity and particle diffusion
coefficients $\nu$ and $D$, which act on vorticity and density fluctuations
respectively, have been introduced, but they could as well be replaced
by hyperdiffusivity terms. An explicit diffusion term $D_{0}\partial_{x}^{2}n_{r}$,
which represents neoclassical diffusion on the radial profile, has
also been added, along with a source term $S_{n}(x,t)$, which represents
particle injection.\\

We could apply the same decomposition procedure to a given radial
profile $u_{r}$ of the $E\times B$ velocity, which would allow us
to introduce the effect of large-scale mean sheared-flows, generated
by the radial electric field $E_{r}$, which plays a fundamental role
in turbulence suppression and pedestal formation in the L-H transition
\citep{biglari:1990,hinton:1993}. However, this is technically more
complicated because the vorticity equation (\ref{eq:hwfd-om}) features
the 1\textsuperscript{st} \emph{and} the 2\textsuperscript{nd} derivatives
of the radial profile of the electrostatic potential $\phi_{r}$ (respectively
the velocity $u_{r}=\partial_{x}\phi_{r}$ and the vorticity $\Omega_{r}=\partial_{x}^{2}\phi_{r}$
radial profiles), and it is not clear which decomposition between
a linear and zonal profile would be the better adapted\footnote{More precisely, would it be better to define $\kappa_{\phi}$ (mean
electrostatic potential gradient), $\kappa_{u}$ (mean $E\times B$
velocity gradient / mean shear) or $\kappa_{\Omega}$ (which would result in a
mean potential vorticity gradient $\kappa_{n}-\kappa_{\Omega}$) ?}. We leave this discussion to a future publication where an external
$E_{r}$ shear acts as an important control parameter.\\
Having presented the theoretical model, we now turn our attention
to the numerical implementation of the FDHW system, using a pseudo-spectral
method with penalisation.

\section{The P-FLARE code: a pseudo-spectral flux-driven code using the penalisation
method}\label{sec:P-FLARE}

In this section, we present the P-FLARE code, which stands for ``Penalised
FLux-driven Algorithm for REduced models'', that we developed to numerically
solve the FDHW equations as its primary example for reduced fluid
model. The code is written in Python using CUDA, and is available
on GitHub \citep{guillon:2025a} under the MIT License.

Pseudo-spectral method provides a fast and precise method for solving
turbulence problems. However, its implementation requires periodic
boundary conditions in all directions in order to compute fast Fourier
transforms. This is not \emph{a priori} satisfied for a flux-driven
system, because we no longer have periodicity along the direction
of the flux drive (\emph{i.e.} the radial, or $x$ direction in our
case). One possible solution that allows us to continue using this
formulation is the method of penalisation \citep{angot:1999,schneider:2005,schneider:2011},
which is based on dividing the radial domain into a physical and a
buffer zone near the edge, where the fluctuations are strongly damped 
and the perturbation parts of the profiles are slightly modified to
make them periodic so that we can still apply the fast Fourier transforms.
In other words, these edge buffer regions act like absorbant walls
for the fluctuations and allow us to impose different kinds of boundary
conditions on radial profiles.

\subsection{Buffer zone}

In the flux-driven formulation, in addition to all the fields being
$L_{y}$ periodic, all non-zonal fluctuations are also $L_{x}$ periodic.
However, this is not the case for example for the zonal density gradient
$\partial_{x}\overline{n}$, which becomes discontinuous at the end
points (orange dots), even though the zonal density perturbation can
be made periodic, as can be seen in figure \ref{fig:decomp}. But,
in order to use the pseudo-spectral algorithm, we need to compute
its fast Fourier transform, since it appears in the nonlinear term
$\partial_{x}\overline{n}\partial_{y}\phi$.

\begin{figure}
\centering{}\includegraphics[width=0.45\textwidth]{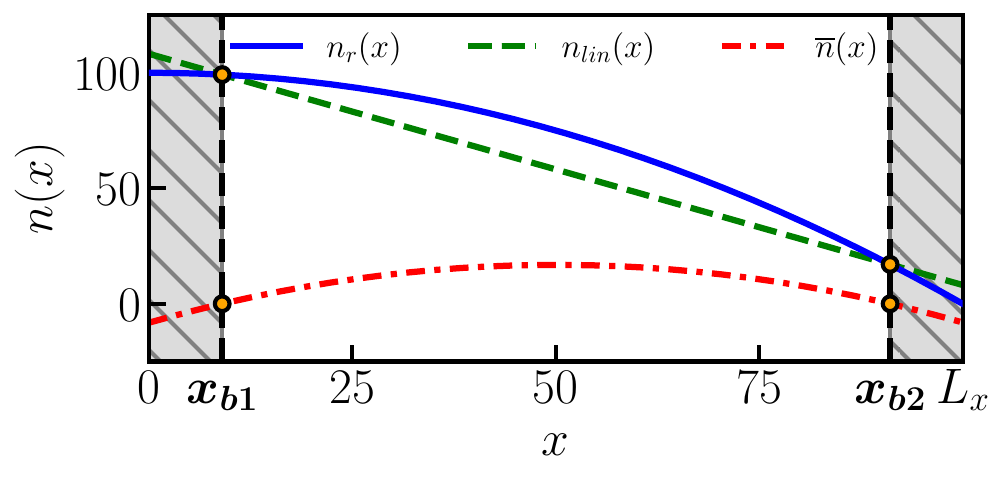}\caption{Decomposition of the radial profile $n_{r}$ (blue) into its linear
component $n_{lin}$ (dashed green) and its zonal fluctuating component
$\overline{n}$ (dashdot red). The background density gradient is
now computed between $x_{b1}$ and $x_{b2}$, and the buffer zones
are indicated by the dashed lines. }\label{fig:decomp-buffer}
\end{figure}

In order to use fast Fourier transforms while maintaining non-periodic
boundary conditions, we need to design a buffer zone at both ends
of the $x$ axis to enforce the desired radial periodicity. For this
purpose, we choose to restrict the radial domain $[0,L_{x}]$ to a
smaller range $[x_{b1},x_{b2}]$, with $0<x_{b1}<x_{b2}<L_{x}$ (see
figure \ref{fig:decomp-buffer}), which describes the physical region
in which the solution is considered valid, and construct buffer zones,
in which the profile is modified slightly in order to get the continuity
of the first few derivatives of the zonal density profile (the continuity
of the second derivative is needed in order to compute the neoclassical
diffusion term).

In the buffer region, the system has no ``physical meaning'' and the
profiles are not realistic, as a result of imposing the perturbation
to match between the two ends of the domain. In order to avoid that
these unrealistic profiles generate spurious transport, we also need
to strongly damp fluctuations in those regions, which is achieved
by the penalisation method.

When decomposing the radial profile, instead of computing the background
density gradient using the endpoints of the domain $x=0$ and $x=L_{x}$,
we now take it between the physical end points $x_{b1}$ and $x_{b2}$
:
\begin{equation}
\kappa(t)\equiv-\frac{n_{r}(x_{b2},t)-n_{r}(x_{b1},t)}{x_{b2}-x_{b1}}\,.\label{eq:kap-buff}
\end{equation}
An example of this decomposition is shown in figure \ref{fig:decomp-buffer}.

Note that $x_{b2}-x_{b1}$ is equal to the physical box size $L_{px}$,
while $L_{x}$, which also includes the two buffer regions, actually
represents the computational domain size.

\subsection{Flattening the zonal density in the buffer}\label{subsec:smoothprof}

The zonal profile \emph{inside} the buffer zone needs to be such that
at least its first two derivatives are continuous at the end points
(\emph{i.e.} $0$ and $L_{x}$ which is the same point in a periodic
box), while keeping the zonal perturbation periodic. This can be achieved
by interpolating the zonal density perturbation in the buffer so that
it basically becomes flat. However, the interpolation is rather costly
from a computational point of view, and therefore a better solution
is to multiply the profile by a ``gate function'', which is exactly
1 outside the buffer zones and 0 at the very edges. In order to ensure
the continuity of the higher derivatives at the ends, the function
that is used should be quite flat when converging to 0 (and $L_{x}$
on the other side), and it should be at least twice differentiable,
to allow the computation of derivatives of the zonal profiles.

\begin{figure}
\centering{}\includegraphics[width=0.8\textwidth]{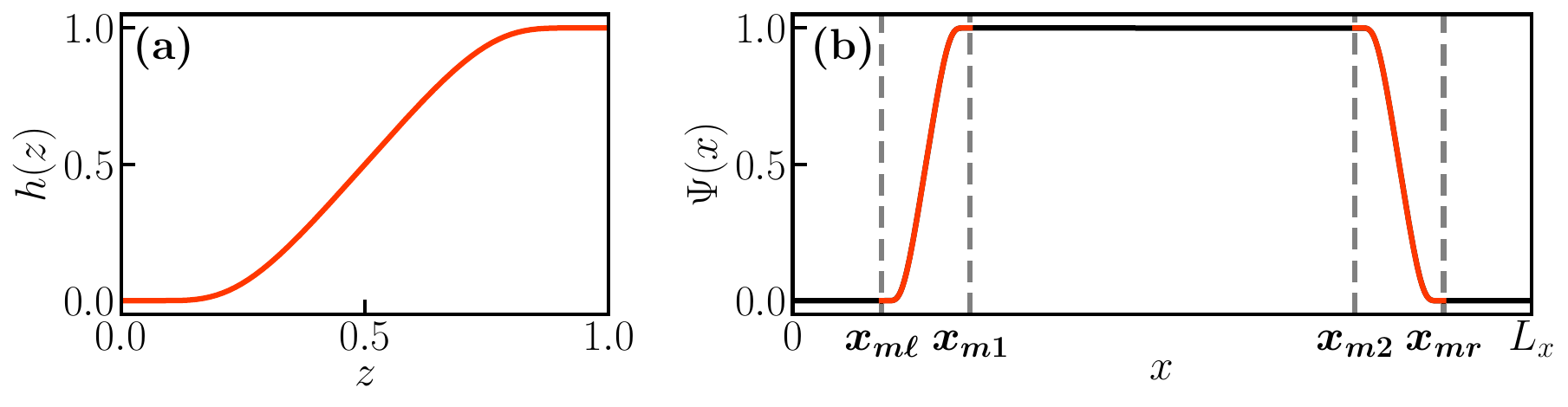}\caption{(a): smooth transition function $h$ defined according to Eq. \ref{eq:smooth-transition}
on $[0,1]$. (b): smooth gate $\Psi$ defined from Eq. \ref{eq:smooth-gate}
on $[0,L_{x}]$, where the transition parts are shown in orange. }\label{fig:smooth-transition}
\end{figure}

In order to guarantee that, we can use \emph{a smooth transition function}
\citep[Ch. 13]{tu:2011} which can be constructed using the following
function, defined for $z\in[0,1]$ as
\begin{equation}
g(z)\equiv\begin{cases}
\exp\left(-1/z\right) & z>0\\
0 & z=0
\end{cases}\,\text{.}\label{eq:expinf}
\end{equation}
This function is infinitely continuous and differentiable in its domain
of validity and all its derivatives are equal to zero at $z=0$, which
makes it rather flat in its vicinity. Using this function as the basis,
we can construct a smooth transition function, defined on $[0,1]$
as
\begin{equation}
h(z)\equiv\frac{g(z)}{g(z)+g(1-z)}\,,\label{eq:smooth-transition}
\end{equation}
which goes from 0 at $z=0$ to 1 at $z=1$, and has all its derivatives
equal to zero at both ends. The function $h$ is illustrated in figure
\ref{fig:smooth-transition} (a).

Scaling and translating this function to match the buffer region,
we can construct the following ``smooth gate'' function $\Psi$:
\begin{equation}
\Psi(x)\equiv\begin{cases}
0 & 0\leq x\leq x_{m_{\ell}}\,,\\
h\left(\frac{x-x_{m\ell}}{x_{m1}-x_{m\ell}}\right) & x_{m\ell}<x<x_{m1}\,,\\
1 & x_{m1}\leq x\leq x_{m2}\,,\\
h\left(\frac{x_{mr}-x}{x_{mr}-x_{m2}}\right) & x_{m2}<x<x_{mr}\,,\\
0 & x_{mr}\leq x\leq L_{x}\,,
\end{cases}\label{eq:smooth-gate}
\end{equation}
 which is exactly 0 in $[0,x_{m\ell}]$ and $[x_{mr},L_{x}]$ (close
to the edges), exactly 1 on $[x_{m1},x_{m2}]$, and which transitions
smoothly for all its derivatives between those intervals. The gate
function is shown in figure \ref{fig:smooth-transition} (b).

\begin{figure}
\centering{}\includegraphics[width=0.7\textwidth]{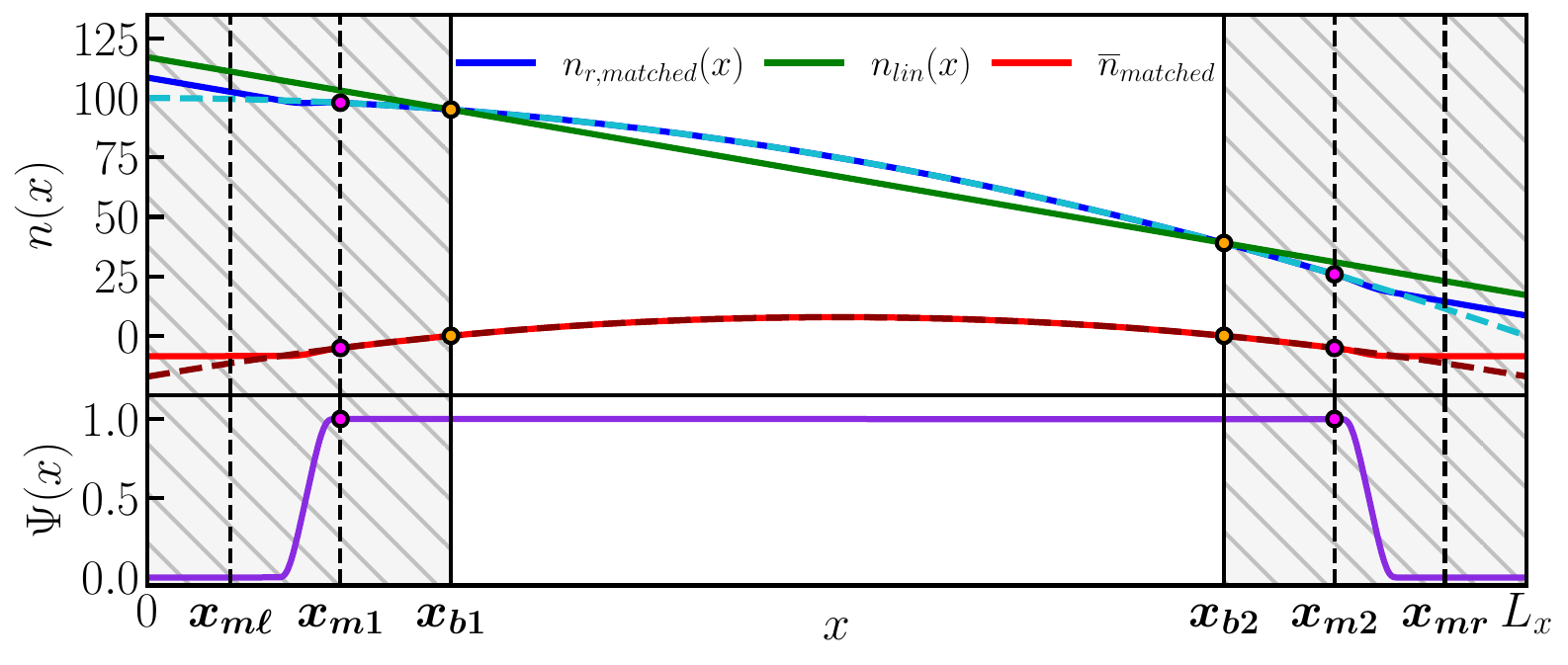}\caption{Top plot: modified zonal density perturbation $\overline{n}_{matched}$
(red), which is now periodic and flat at both ends. The corresponding
matched radial profile $n_{r,matched}$ is shown in blue. The orignal
radial profile $n_{r}$ (dashed lightblue), linear profile $n_{lin}$
(green) and zonal profile $\overline{n}$ (dashed darkred) from figure
\ref{fig:decomp-buffer} are also shown. Bottom plot: smooth-gate
function $\Psi$.}\label{fig:decomp-smooth}
\end{figure}

Notice that the intervals around the endpoints where the gate function
is exactly zero are introduced for numerical reasons: since the $x$
axis is discretised in numerical simulations, it is better to ensure
that the derivatives are exactly zero on a small portion of the $x$
grid, rather than just at the very ends. These regions $[0,x_{m\ell}]$
and $[x_{mr},L_{x}]$ are exagerated in figures \ref{fig:smooth-transition}
and \ref{fig:decomp-smooth} for visual purposes. In practice, we
will choose $x_{m\ell}$ and $x_{mr}$ such that they correspond to
some of the very first and last points of the grid (obviously within
the buffer region) respectively (see section \ref{subsec:Spread-setup}
for an actual application). In the following, we actually use the
radial extension of the transition
\begin{equation}
\delta x_{m}\equiv x_{m1}-x_{m\ell}=x_{mr}-x_{m2}\label{eq:deltaxm}
\end{equation}
 as a parameter, and infer the individual values of $x_{m\ell}$ and
$x_{mr}$ from that. \\

Now, we can use the gate function $\Psi$ to modify the zonal density
profile inside the buffer zones and match its values and derivatives
at both edges. The simplest solution would be to just multiply the
zonal density perturbation by $\Psi$ in order to ensure the periodicity
of $\overline{n}$ and all its derivatives. However, the values of
the density perturbation, as computed from the decomposition, can
be very different from zero at $x=0$ and $x=L_{x}$, and forcing
them to be zero could introduce strong gradients in the buffer zones
which might trigger numerical instabilities (even though they would
be suppressed in the buffer). In order to mitigate that effect, we
first shift the zonal perturbation by an offset: 
\begin{equation}
n_{off}\equiv\frac{1}{4}\left[\overline{n}(0)+\overline{n}(x_{m1})+\overline{n}(x_{m2})+\overline{n}(L_{x})\right]\,,\label{eq:nbaroff}
\end{equation}
corresponding to the mean value between both ends of the profile and
its value at $x=x_{m1}$ and $x=x_{m2}$. Then, we apply the gate
function, and shift back the profile by $n_{off}$:
\begin{equation}
\overline{n}_{matched}\equiv\Psi(x)\left(\overline{n}-n_{off}\right)+n_{off}\,.\label{eq:nbarsmth}
\end{equation}
Note that this does not change the profile within the interval $[x_{m1},x_{m2}]$.
The matched zonal perturbation profile is shown on figure \ref{fig:decomp-smooth}
(red line). We also show the corresponding ``matched'' radial profile
(blue line), which is the combination of the matched zonal perturbation
and the linear profile
\begin{equation}
n_{r,matched}\equiv n_{lin}+\overline{n}_{matched}\,,\label{eq:nrsmth}
\end{equation}
actually ``seen'' by the system.

Notice that we initially took $\left(x_{m1},x_{m2}\right)=\left(x_{b1},x_{b2}\right)$,
but this introduced strong density gradients very close to the physical
space $[x_{b1},x_{b2}]$, where the fluctuations are not completely
zero. Instead, taking $x_{m1}$ and $x_{m2}$ well inside the buffer
zone allows us to avoid such effects.

In practice, we subtract the mean value $n_{m}=\langle\overline{n}_{matched}\rangle_{x}$
from the matched zonal profile, as already discussed in section \ref{subsec:eqs-FD}
(see Eq. \ref{eq:zonal_0mean}).

\subsection{Penalisation method}

Having constructed a profile whose zonal perturbation is periodic
and whose derivative falls to zero at the endpoints, we now want to
impose some constraints on both the radial profiles and turbulent
fluctuations inside the buffer zone, in order to enforce the desired
boundary conditions and avoid spurious fluctuations inside the edge
regions. To this end, we use the volume penalisation approach, introduced
by Ref. \citealp{angot:1999} for Navier-Stokes simulations with a
bounded, non-periodic domain, and used later for pseudo-spectral simulations
of hydrodynamic \citep{schneider:2005} and MHD \citep{schneider:2011}
turbulence.

The general idea is to impose a strong friction term
\begin{equation}
-\mu H(x)(A-A_{buff})\label{eq:pendef}
\end{equation}
to force the field $A$ to rapidly decay to a given boundary function
$A_{buff}$ inside the buffer zone, represented by the mask function
$H$, which is 0 in the physical domain, and 1 inside the buffer.
In order to underline that it is large, the friction coefficient $\mu$
is sometimes represented as $1/\varepsilon$, where $\varepsilon$
stands for some permeability coefficient, for the case when $A$ is
the velocity field. This corresponds to building a thin artificial
boundary layer around the buffer zone, in which the fields are forced
to rapidly converge towards their edge value.

Using (\ref{eq:pendef}) and the Eqs. \ref{eq:ntotbis} and (\ref{eq:hwfd-om}),
we can write the penalised form of the equations that we solve numerically:
\begin{subequations}
\begin{align}
\partial_{t}\Omega+[\phi,\Omega] & =C(\widetilde{\phi}-\widetilde{n})+\nu\nabla^{2}\widetilde{\Omega}-\mu\nabla\times\left(H(x)\mathbf{v}_{E\times B}\right)\,,\label{eq:hw-om-pen}\\
\partial_{t}N+[\phi,N] & =C(\widetilde{\phi}-\widetilde{n})+D\nabla^{2}\widetilde{n}+S_{n}(x,t)-\mu H(x)\left(N-n_{r,buff}\right)\,,\label{eq:hw-ntot-pen}
\end{align}
\end{subequations}
where the last two terms in both equations correspond to the penalisation.
Notice that in Eq. \ref{eq:hw-om-pen}, we impose the fluid velocity
$\mathbf{v}_{E\times B}=-\partial_{y}\phi\mathbf{e}_{x}+\partial_{x}\phi\mathbf{e}_{y}$
to be zero inside the buffer zone, corresponding to \emph{no-slip}
boundary conditions, and since the equation is actually for the vorticity
$\Omega=\nabla\times\mathbf{v}_{E\times B}$, it is the \emph{curl}
of that penalisation condition (projected onto $\mathbf{e}_{z}$)
that appears in this equation \citep{schneider:2005,schneider:2011}.
For the density equation (\ref{eq:hw-ntot-pen}), the total density
$N$ is forced to be equal to a given boundary function $n_{r,buff}(x,t)$
in the buffer, which is only a function of $x$ and time, so that
the turbulent, non-zonal fluctuations are forced to be zero in the
buffer. The choice for $n_{r,buff}$ will be discussed further in
section \ref{subsec:BC-radial} about boundary conditions on the radial
profile.

We can thus write the penalised equations for the turbulent fluctuations
(using Eqs. \ref{eq:hwfd-om} and \ref{eq:hwfd-nturb})
\begin{subequations}
\begin{align}
\partial_{t}\widetilde{\Omega} & +[\phi,\widetilde{\Omega}]-\langle[\phi,\widetilde{\Omega}]\rangle_{y}=C(\widetilde{\phi}-\widetilde{n})+\nu\nabla^{2}\widetilde{\Omega}-\mu\nabla\times\left(H(x)\hat{z}\times\nabla\widetilde{\phi}\right)\,,\label{eq:hw-omturb-pen}\\
\partial_{t}\widetilde{n} & +\left(\kappa(t)-\partial_{x}\overline{n}\right)\partial_{y}\widetilde{\phi}+\partial_{x}\overline{\phi}\partial_{y}\widetilde{n}+\widetilde{[\phi,n]}=C(\widetilde{\phi}-\widetilde{n})+D\nabla^{2}\widetilde{n}-\mu H(x)\widetilde{n}\,,\label{eq:hw-nturb-pen}
\end{align}
\end{subequations}
and for the radial profile of the velocity $u_{r}$, which is just
the zonal velocity $\overline{v}_{y}$ here\footnote{Although the theoretical model in section \ref{subsec:eqs-FD} had
an equation on the total vorticity $\Omega=\overline{\Omega}+\widetilde{\Omega}$,
we prefer to separate both components in the implementation. Furthermore,
we write an equation for $\partial_{t}\overline{v}_{y}$ instead of
$\partial_{t}\overline{\Omega}$, since it makes more sense to think
in terms of the radial profile of the $E\times B$ velocity, rather
than that of vorticity. Moreover, this paves the way for a more general
version of the model which could include a non-zero background velocity
profile $u_{r}$, or damping towards a background neoclassical flow.} and the radial density $n_{r}$
\begin{subequations}
\begin{align}
\partial_{t}\overline{v}_{y} & =\langle\widetilde{\Omega}\partial_{y}\widetilde{\phi}\rangle_{y}-\mu H(x)\overline{v}_{y}\label{eq:hw-ubar-pen}\\
\partial_{t}n_{r} & =\partial_{x}\langle\widetilde{n}\partial_{y}\widetilde{\phi}\rangle_{y}+S_{n}(x,t)+D_{0}\partial_{x}^{2}n_{r}-\mu H(x)\left(n_{r}-n_{r,buff}\right)\,.\label{eq:hw-nr-pen}
\end{align}
\end{subequations}

\begin{figure}
\centering{}\includegraphics[width=0.49\textwidth]{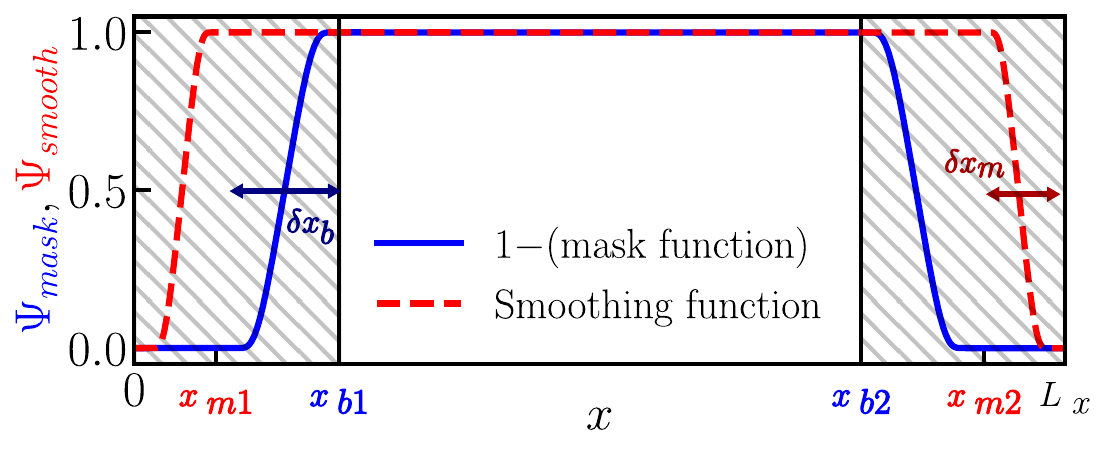}\caption{Combined plot of $\psi_{mask}(x)=1-H(x)$ (blue) and the smooth-gate
function $\Psi_{smooth}(x)$ (dashed red). Their radial extensions
$\delta x_{b}$ and $\delta x_{m}$ are also shown (respectively blue
and red arrows).}\label{fig:pen-func}
\end{figure}

The mask function $H(x)$, which should quickly (but smoothly) transition
from 0 to 1 as we enter the buffer zone, is taken as the smooth gate
function $\Psi$ defined previously in Eq. \ref{eq:smooth-gate}.
More precisely, we will use the following mask function:
\begin{equation}
H(x)=1-\Psi_{mask}(x)\,.\label{eq:maskpen}
\end{equation}
Note that $\Psi_{mask}$ is parametrised differently from the function
used to make zonal profiles periodic in section \ref{subsec:smoothprof}.
As illustrated on figure \ref{fig:pen-func}, the parameters are chosen
so that fluctuations are zero before the radial profile has any modification.
In other words the fluctuations do not ``see'' the modified parts
of the radial profiles.

\subsection{Boundary conditions}\label{subsec:BC-radial}

\subsubsection{Boundary profile in the buffer zone}

First, we specify the boundary radial profile $n_{r,buff}(x,t)$ that
we impose in the buffer zone. Since the fluctuations are not supposed
to penetrate that region, we argue that the density profile should
keep its initial shape $n_{r}(x,t=0)$. However, since turbulence
creates a net outward radial particle flux, the total number of particle
can vary in time in the physical domain. This would create large gradients
in the buffer zone if we kept the density in the buffer region really
constant. In order to avoid this, we need the boundary profile to
decrease or increase together with the value of $n_{r}(x,t)$ at $x_{b1}$
and $x_{b2}$, where the buffers meet the physical domain. In order
to achieve this, we propose the following boundary profile:
\begin{equation}
n_{r,buff}(x,t)=\begin{cases}
n_{r}(x,t=0)-n_{r}(x_{b1},t=0)+n_{r}(x_{b1},t) & x\leq x_{b1}\\
n_{r}(x,t=0)-n_{r}(x_{b2},t=0)+n_{r}(x_{b2},t) & x\geq x_{b2}
\end{cases}\,,\label{eq:boundary-nr}
\end{equation}
which will cause the boundary profiles to move together with $n_{r}(x_{b1},t)$
and $n_{r}(x_{b2},t)$ at these points, without changing their initial
shape.\\

\subsubsection{Boundary condition at $(x_{b1},x_{b2})$}

As a first application of the code with non-trivial boundary conditions,
we choose to fix the density profile at the right buffer, $x=x_{b2}$
(Dirichlet boundary condition), while letting it evolve freely at
the left buffer, $x=x_{b1}$, as a response to the turbulent particle
flux (which is a loosely Neumann boundary condition). Note that we
make this somewhat arbitrary choice because we are interested in studying
the profile relaxation, or its evolution with a particle source localised
at the inner boundary, which is more relevant if the profile is fixed
at the outer boundary. It also allows us to demonstrate the two kinds
of boundary conditions that can be used.

In particular, imposing the Dirichlet boundary condition $\partial_{t}n_{r}(x_{b2},t)=0$
is not straightforward: even though we impose a boundary profile for
$n_{r}$ inside the buffer, the condition is only really met where
the mask function $H(x)$ is close to $1$. Therefore, in the vicinity
of $x_{b2}$ in the buffer, where $H(x)\ll1$ (see figure \ref{fig:pen-func}),
the radial profile can still evolve freely because the effect of penalisation
is quite weak in this region. In order to impose the condition exactly
at $x_{b2}$, but in a smooth way around this point, we apply an artificial
Gaussian sink centered at the outer boundary point. Indeed, we cannot
impose a boundary condition only at $x_{b2}$, because such a singularity
would trigger numerical errors when computing fast Fourier transforms.

This artificial Gaussian sink is taken as
\begin{equation}
S_{b2}(x,t)=-\partial_{t}n_{r}(x_{b2},t)\exp\left[-(x-x_{b2})^{2}/(2\sigma_{S_{b}}^{2})\right]\,,\label{eq:sourceartif2}
\end{equation}
where $\partial_{t}n_{r}(x_{b2},t)$ is computed from Eq. \ref{eq:hw-nr-pen}.
The sink is centered on $x_{b2}$ and imposes $\partial_{t}n_{r}(x_{b2},t)=0$.
Since it has some finite (small) width $\sigma_{S_{b}}$, its effect
extends into the physical space around $x_{b2}$, which means that
it modifies the particle budget inside the physical domain and acts
as a volumetric particle sink. The particle budget becomes:
\begin{equation}
d_{t}N_{\varphi}=\int_{x_{b1}}^{x_{b2}}\partial_{t}n_{r}\,dx=\Gamma_{n}(x_{b1},t)-\Gamma_{n}(x_{b2},t)+P_{b2}(t)\,,\label{eq:Nconserv}
\end{equation}
with $N_{\varphi}$ the total density inside the physical domain $[x_{b1},x_{b2}]$,
$\Gamma_{n}$ the total particle flux (turbulent and potentially diffusive
if the neoclassical diffusion $D_{0}$ is also included) and $P_{b2}=\int_{x_{b1}}^{x_{b2}}S_{b2}(x,t)\,dx$
is the integral over the physical domain of the sink $S_{b2}$ localised
at $x_{b2}$.

Note that this is a particular choice for the present study, and that
the boundary conditions can easily be modified for other applications
of the code.

\subsection{Limits of the penalisation method}\label{subsec:limitpen}

First of all, notice that, by construction, turbulent (zonal and non-zonal)
perturbations are taken to be zero mean-valued. But since we impose
the radial profile of the poloidal velocity $u_{r}$ to be zero in
the buffer zone, it may acquire a non-zero mean value resulting in
$\overline{v}_{y}=u_{r}-\langle u_{r}\rangle_{x}\neq u_{r}$. We account
for this small non-zero mean value in the numerical implementation,
which can be seen as a Doppler shift (and which remains several orders
of magnitude lower than the typical amplitude of $u_{r}$ in the saturated
state). This makes it so that the zonal velocity perturbation is not
exactly zero in the buffer zone, but it is actually equal to $-\langle u_{r}\rangle_{x}$.
This is not really a problem, but a feature of the penalisation with
the particular kind of boundary condition that is used.

Second, the actual effectiveness of the penalisation is related to
the value of the penalisation coefficient $\mu$ in such a way that
the higher the coefficient $\mu$, the closer we are to a system with
actual impermeable boundaries. But since $\mu<\infty$, there is still
some permeability which means that some fluctuations manage to penetrate
into the buffer region. Nevertheless, we observe in simulations that
the turbulence level in the buffer zone is still 2 to 3 orders of
magnitude below that of what one finds in the physical domain, and
no coherent information seems to pass from one buffer to the other.
This can in principle be improved by increasing the numerical value
of the penalisation coefficient, but experience shows that the improvement
remains limited, and it causes the adaptive time stepping algorithm
that we use to slow down considerably.

The mask function can also create a second instability induced by
the penalisation term $-\mu\nabla\times\left(H(x)\hat{z}\times\nabla\widetilde{\phi}\right)$
from Eq. \ref{eq:hw-omturb-pen}. Indeed this term can be decomposed
into a penalisation term on the vorticity $-\mu H(x)\nabla_{\perp}^{2}\widetilde{\phi}$,
and a second term $-\mu\nabla H\times\left(\hat{z}\times\nabla\widetilde{\phi}\right)$
which features the gradient of the mask function and which will appear
as an additional term in the dissipative drift-wave dispersion relation
when kept, in contrast to Navier-Stokes, where there is no linear
instability. Note that this is not a numerical artefact \emph{per
se}: this would probably actually happen if we had actual walls around
the system as implied by the penalisation scheme. However it is unwelcome
here because the buffer zones are not supposed to represent actual
walls, and it shows that using the penalisation method may cause the
dynamics close to the boundaries to be affected in specific ways,
which may be different from what is intended. Fortunately, this instability
remains of very low amplitude, confined to the vicinity of the buffer
zone, and is completely suppressed once ``real'' turbulence reaches
the buffer and covers it (see the discussion in section \ref{subsec:spreadingDNS}
on an actual DNS). Again this effect can be reduced by making the
buffer regions larger, so that the local gradients of the mask function
become smaller. However, this is a trade-off, since in doing so, one
looses a bigger part of the computational domain to buffers.

\section{Turbulent spreading and transport}\label{sec:spread_transport}

In order to demonstrate the capabilities of our code, we consider
the decades-old problem of turbulent spreading coupled to turbulent
transport. While the exact definition of spreading remains controversial,
a rather strong definition related to observations of turbulence in
linearly stable regions \citep{guo:2017,heinonen:2020a} can be ruled
out in our model. However a weaker definition consists of the nonlinear
tendency of turbulence to spread itself, which can be represented
by a nonlinear diffusive term in the equation for the fluctuations.
Such a term is well known to not be very efficient in penetrating
stable or damped regions, but modifies the complex feedback loops
between turbulence and transport in such a way that it increases the
initial nonlinear diffusion of turbulence and the profile together
into an initially stable region. Note, however, that the dissipative
drift instability that we consider here does not have an instability
threshold in the inviscid limit.

Our code allows us to study the detailed dynamics of how turbulent
fluctuations, due to an initial density profile with a highly local
instability drive, spreads in the radial direction, when the system
is released with no forcing. Naturally, this happens through a coupled
evolution of the profile and turbulence which results in a turbulent
front propagation accompanied by a relaxation of the initial density
profile.

For this purpose, we construct an initial profile which is very steep
on the left, and then very flat on the right using the following expression:
\begin{equation}
n_{r}(x,t=0)=\frac{L_{x}}{\alpha}\left[\tanh\left(\frac{x_{a}-x}{L_{x}}\kappa_{\ell}\alpha\right)-\tanh\left(\frac{x_{a}-L_{x}}{L_{x}}\kappa_{\ell}\alpha\right)\right]\,,\label{eq:nr0_spread}
\end{equation}
where $\kappa_{\ell}$ is the slope of the steepest part (\emph{i.e.}
the background density gradient on the left), $x_{a}$ defines the
center of this region, and $\alpha$ is a scaling factor that controls
the width and the height of the hyperbolic tangent. The second term
is an offset which allows us to have $n_{r}(x=L_{x},t=0)=0$. Note
that, for $x\sim x_{a}$, we have $n_{r}(x,t=0)\sim-\kappa_{\ell}(x-x_{a})$,
which means that $\kappa_{\ell}$ indeed corresponds to the local
background density gradient at that point. The linear instability
is also the strongest at this point, where the profile is the steepest,
and naturally the turbulence initially develops there. The turbulent
patch then spreads to the right, accompanied by a relaxing profile
until they reach the right buffer zone. Note that while a relaxing
profile would drag the turbulence with it, resulting in a similar
overall picture, the existence of turbulence spreading, in the form
of a flux term in the fluctuation evolution, changes the details of
how exactly this happens.

\begin{figure}
\centering{}\includegraphics[width=0.95\textwidth]{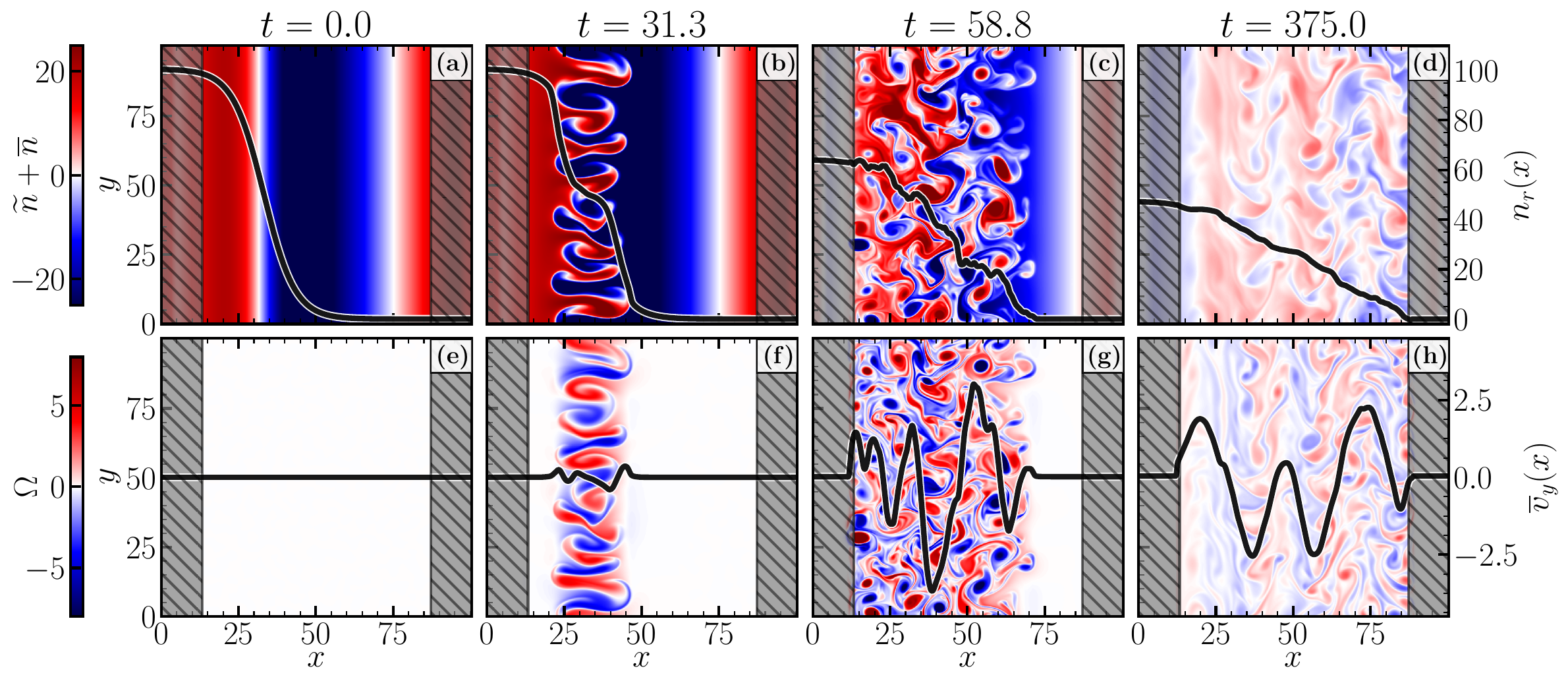}\caption{Spreading of the turbulent region in a numerical simulation of padded
resolution $1024\times1024$, with $C=0.05$ and $\kappa_{\ell}=5$.
Top plots (a-d): snapshots of the density fluctuation $n(x,y)$ and
radial density profile $n_{r}$ (black line). Bottom plots (e-h):
snapshots of the vorticity fluctuation $\Omega(x,y)$ and zonal velocity
profile $\overline{v}_{y}$ (black line). Buffer zones are shown in
dashed lines. }\label{fig:snapshots-spread}
\end{figure}

The evolution of density and vorticity fluctuations as well as the
zonal velocity and the density profiles are shown in figure \ref{fig:snapshots-spread},
for a numerical simulation of padded resolution $1024\times1024$,
with $C=0.05$, $\kappa_{\ell}=5$, $L_{x}=L_{y}=32\pi$, $x_{a}=2.5x_{b1}$,
and $\alpha=2$. The animation for this simulation is available on
the repository page \citep[ Movie 1]{guillon:2025b}. We can describe
the time evolution of the system as follows:
\begin{enumerate}
\item The initial radial density profile at $t=0$ (a) is the steepest at
$x=x_{a}\approx33$, so the linear instability mostly occurs around
this point. Initially, there are no zonal flows (e).
\item At $t=31.3$, the instability is well developed, and we see a local
flattening (b) of the profile centered on $x=x_{a}$. The 2D map of
density exhibits blobs and holes of positive and negative density
fluctuations respectively, alternating periodically in the poloidal
direction. Positive blobs move to the right while negative holes move
to the left. For the velocity profile (f), this corresponds to the
formation of a nonlinear structure similar to a modulational instability.
The vorticity map exhibits similar structures.
\item At $t=58.8$, as shown in plots (c) and (g), the instability has completely
saturated and the left part of the domain is in a turbulent state.
The initial density profile has collapsed, due to the outward turbulent
particle flux, and has relaxed to the right.
\item At $t=375$, plots (d) and (h), the turbulent region has completely
spread to the right of the radial axis. The global density profile
is considerably flatter than the initial one, and the global $C/\kappa$
is thus higher. As a result, the system has made the transition to
the zonal flow dominated state, which is evidenced by the large scale
sheared flows (h) and the density corrugations (d), where the so-called
staircase steps can be observed. The zonal flows reduce the particle
flux, hence the profile stops collapsing and the system ends up in
some kind of marginal state, that is mostly ``frozen'' by the zonal
flows. The turbulence level is also smaller, as shown in the 2D maps
of density and vorticity.
\end{enumerate}
In the following, we study these different stages in detail: (ii) the
``pair formation'' in section \ref{subsec:early-time}, (iii) the spreading
of the turbulent region, along with the relaxation of the density
profile in section \ref{subsec:Spreading}, which we reproduce using
a 1D reduced model, and finally (iv) the formation of zonal flows once
the density profile is sufficiently flat at the end, along with the
``freezing'' of the system due to zonal flow domination, in section
\ref{subsec:Latest-times}.

We have also made a detailed comparison to the BOUT++ code, which can be considered as the standard in finite difference formulation where the boundary conditions can be specified explicitly, and the results
of this benchmark, which shows reasonable agreement, can be found in Appendix \ref{sec:Benchmark}.

\subsection{Numerical set-up}\label{subsec:Spread-setup}

To further study the dynamics of the flux-driven Hasegawa-Wakatani
system accurately, we perform high resolution ($4096\times4096$)
numerical simulations, for which the physics and penalisation parameters
are given in table \ref{tab:spread_params}. The parameters that determine
the buffer and smoothing functions correspond to integer multiples
of the unpadded grid resolution $\Delta x=L_{x}/N_{x}$, where $N_{x}=2\lfloor4096/3\rfloor=2730$.
The initial value for the profile is given by Eq. \ref{eq:nr0_spread}
as before.

\begin{table}
\begin{centering}
\begin{tabular*}{0.99\textwidth}{@{\extracolsep{\fill}}>{\centering}m{0.12\textwidth}>{\centering}m{0.1\textwidth}>{\centering}m{0.12\textwidth}>{\centering}m{0.12\textwidth}>{\centering}m{0.08\textwidth}>{\centering}m{0.08\textwidth}>{\centering}m{0.18\textwidth}}
\toprule
\multicolumn{4}{c}{Physical parameters} & \multicolumn{3}{c}{Initial profile} \vspace{0.2cm}\tabularnewline
$L_{x},L_{y}$ & $C$ & $\nu=D$ & $D_{0}$ & $\kappa_{\ell}$ & $\alpha$ & $x_{a}$\vspace{0.2cm}\tabularnewline

$64\pi$, $64\pi$ & $0.05$ & $6.6\times10^{-3}$ & $0$ & $5$ & $2$ & $8x_{b1}=53.04$\tabularnewline
\midrule
\end{tabular*}
\begin{tabular*}{0.99\textwidth}{@{\extracolsep{\fill}}>{\centering}p{0.06\textwidth}>{\centering}p{0.06\textwidth}>{\centering}p{0.06\textwidth}>{\centering}p{0.06\textwidth}>{\centering}p{0.06\textwidth}>{\centering}p{0.06\textwidth}>{\centering}p{0.2\textwidth}>{\centering}p{0.2\textwidth}}
\multicolumn{8}{c}{Penalisation parameters}\vspace{0.2cm}\tabularnewline

\multicolumn{3}{c}{Buffer one} & \multicolumn{3}{c}{Smoothing function} & Penalisation coefficient & Artificial source size\vspace{0.2cm}\tabularnewline
$x_{b1}$ & $x_{b2}$ & $\delta x_{b}$ & $x_{m1}$ & $x_{m2}$ & $\delta x_{m}$ & $\mu$ & $\sigma_{S_{b}}$\vspace{0.2cm}\tabularnewline

6.63 & 194.43 & 4.42 & 3.31 & 197.75 & 2.95 & 100 & $5L_{x}/N_{x}=0.37$\tabularnewline
\bottomrule
\end{tabular*}\caption{Physics and penalisation parameters for the $4096\times4096$ padded
simulation. The initial profile is taken as Eq. \ref{eq:nr0_spread}.
}\label{tab:spread_params}
\par\end{centering}
\end{table}

Note that for this high resolution simulation, the buffer zone corresponds
to less than $7\%$ of the total computational domain size. In contrast,
in the $1024\times1024$ simulation shown in figure \ref{fig:snapshots-spread},
the buffer zone covers $25\%$ of the computational domain size (using
the same integer multiples for setting the buffer zone).

The time evolution of energy, flux and density at the left boundary
$n_{r}(x_{b1},t)$ are shown in figure \ref{fig:0Drelax} in Appendix
\ref{sec:0Drelax}.

\subsection{Early times: Pair production}\label{subsec:early-time}

First, we detail the flattening of the initial profile as a result
of positive blobs of density $\delta n_{+}$ travelling outwards,
and negative holes $\delta n_{-}$ travelling inwards. This can be
seen as a consequence of the modulation of the linear drift-wave instability,
which occurs mostly where the profile is the steepest (highest local
$\kappa$). Although the waveform of this instability is mainly poloidal
(\emph{i.e.} $k_{y}=0$ modes have zero growth rate), the nonlinear
interaction of two linearly unstable modes can form a structure with
a finite radial scale through a mechanism akin to modulational instability.

\begin{figure}
\centering{}\includegraphics[width=0.85\textwidth]{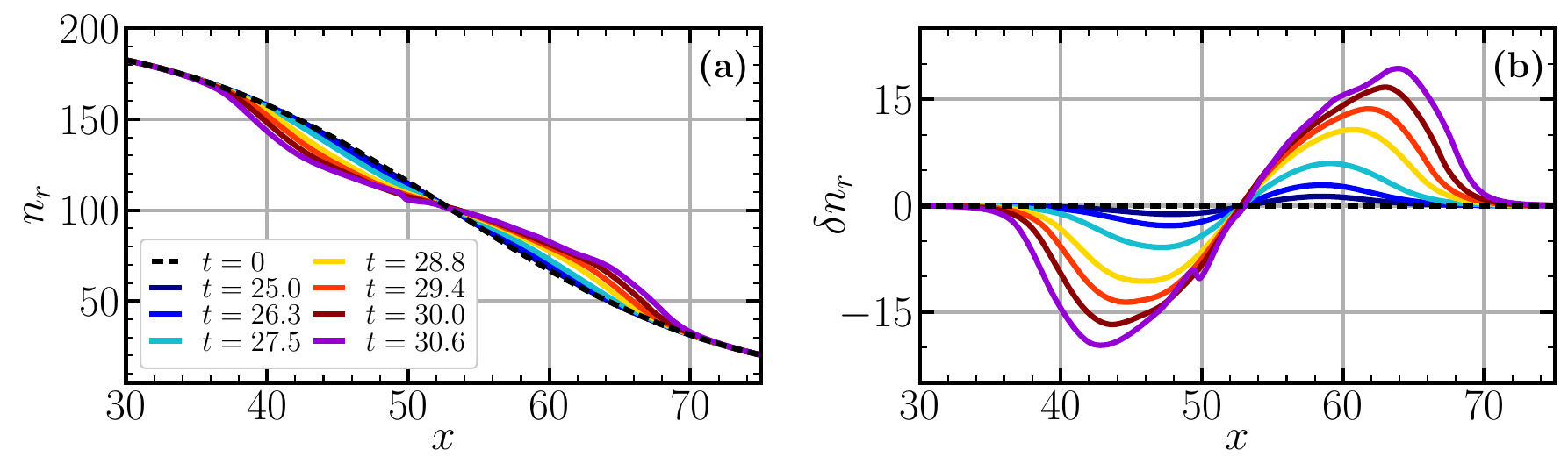}\caption{Flattening of the initial density profile in the early times. Left
plot (a): density profile restrained on the interval $x\in[30,75]$
at different time steps. The initial profile is in black dashed lines.
Right plot (b): perturbation $\delta n_{r}=n_{r}-n_{r}(t=0)$ from
the intial profile.}\label{fig:avalanche_times}
\end{figure}

The evolution of the profile is shown in figure \ref{fig:avalanche_times}
(a) at different time steps (the initial profile is the black dashed
line). The flattening is centered at $x=x_{a}\approx53$, around which
the modulation grows, making the profile more steep at the edges.
The perturbation with respect to the initial profile 
\begin{equation}
\delta n_{r}\equiv n_{r}-n_{r}(t=0)\label{eq:deltanr}
\end{equation}
is shown in figure \ref{fig:avalanche_times} (b), where we clearly
see the formation of a negative ``dip'' on the left, corresponding
to the holes $\delta n_{-}$, and a positive bump on the right, corresponding
to the positive blobs $\delta n_{+}$, both moving away from $x=x_{a}$
and, as a result, flattening the profile. The growth of this flattening
perturbation, which we may call an avalanche from the point of view
of the profile, is initially smooth, but becomes chaotic towards the
end $t=30.6$ (violet line). At this point, the system really enters
the turbulent regime, as the blobs become unstable themselves, forming
smaller scale structures, as can be seen in plots (c) and (g) of figure
\ref{fig:snapshots-spread}.

To quantify this dynamics, we can look at the time evolution of the
root mean square (rms) of the deviation from the initial profile 
\begin{equation}
\langle\delta n_{r}\rangle_{rms}\equiv\left(\frac{1}{x_{2}-x_{1}}\int_{x_{1}}^{x_{2}}\delta n_{r}^{2}\,dx\right)^{1/2}\,,\label{eq:deltanrrms}
\end{equation}
where we take $x_{1}=30$ and $x_{2}=75$ since it is sufficient to
look only in the vicinity of the avalanche. We can also look at the
position $X_{\pm}$ of the positive and negative perturbations $\delta n_{\pm}$
that we observe on figure \ref{fig:avalanche_times} (b). We define
those as the barycenter of the squared perturbations:
\begin{equation}
X_{\pm}(t)\equiv\frac{\int_{x_{1}}^{x_{2}}x\delta n_{r}^{2}\Theta\left(\pm\delta n_{r}\right)\,dx}{\int_{x_{1}}^{x_{2}}\delta n_{r}^{2}\Theta\left(\pm\delta n_{r}\right)\,dx}\,,\label{eq:Xblobs}
\end{equation}
where $\Theta$ is the Heaviside step function. This definition is
similar to tracking the position of the maxima, but it has the advantage
of behaving more smoothly.

\begin{figure}
\centering{}\includegraphics[width=0.95\textwidth]{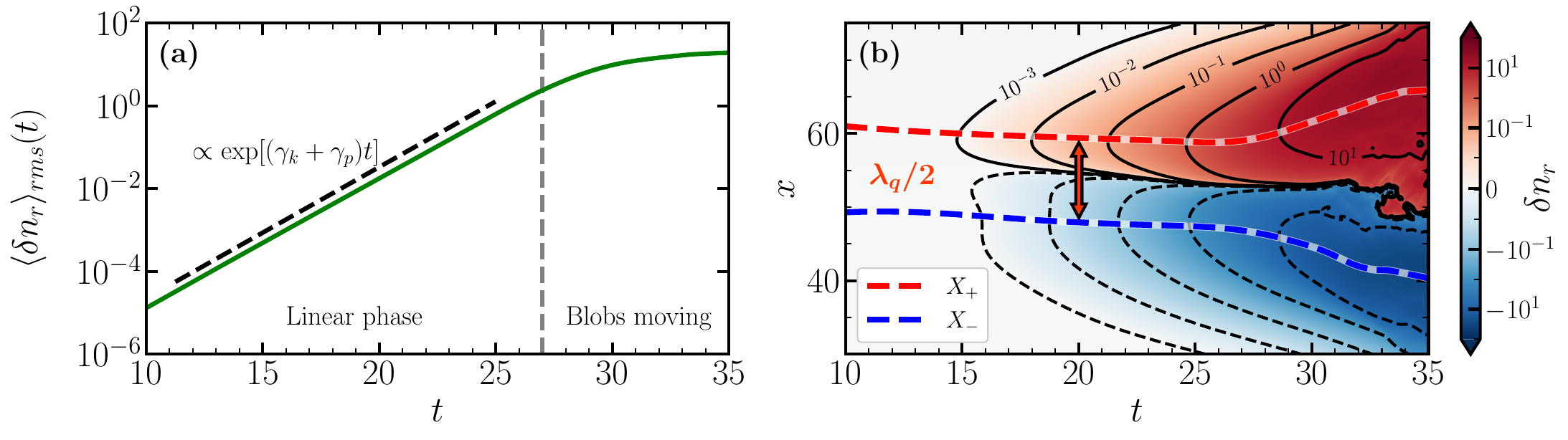}\caption{Left plot (a): time evolution of the rms value $\langle\delta n_{r}\rangle$
of the density profile perturbation. The slope of the growth of the
perturbation in the linear phase is compared to the sum of the growth
rates of the most unstable mode and side-band $\gamma_{k}+\gamma_{p}$.
Right plot (b): positions $X_{\pm}$ of the density perturbations
$\delta n_{\pm}$ as functions of time, shown respectively in red
and blue dashed lines. The contour plots of the density perturbation
$\delta n_{r}$ are also shown. The perturbation wavelength $\lambda_{q}=2(X_{+}-X_{-})\approx23.0$
is estimated at $t\approx20$ (orange arrow).}\label{fig:avalanche_growth}
\end{figure}

In figure \ref{fig:avalanche_growth}, we show the rms value $\langle\delta n_{r}\rangle_{rms}$
of the perturbation (a) and the position $X_{\pm}$ of the bump and
the hole (b) as functions of time. The time evolution of the perturbation
can be decomposed in two phases. First, there is a linear phase, where
the perturbation grows exponentially and the bump and the hole stay
almost at the same distance from each other, although we can notice
that they slowly drift towards the left (in other simulations they
may not move at all). Then, for $t\geq27$, the amplitude of the perturbation
saturates, which is associated to the positive perturbation moving
radially outward while the negative perturbation moves inward, at
roughly constant speeds. Finally, towards the end, the perturbation
starts to be chaotic and the blobs become unstable themselves.

To explain the formation of the perturbation, which is a radial structure,
we can assume that it results from triadic interactions between the
most unstable mode $k=(0,k_{y0})$, a radial mode $q=(q,0)$, with
$q$ corresponding to the wavenumber of the perturbation, and two
side-bands $p_{\pm}=(\pm q,k_{Y0})$, which statisfy the triadic interaction
$\mathbf{k}-\mathbf{p}_{\pm}\pm\mathbf{q}=\mathbf{0}$. Taking the
Fourier transform of Eq. \ref{eq:hw-nr-raw}, and restraining the
nonlinear term to only those, yields
\begin{equation}
\partial_{t}\overline{n}_{q}=qk\left(\phi_{k}^{*}n_{p_{+}}-\phi_{p_{+}}n_{k}^{*}-\phi_{k}n_{p_{-}}^{*}+\phi_{p_{-}}^{*}n_{k}\right)\,.\label{eq:dtnqzonal}
\end{equation}
Assuming $\phi_{k,p_{\pm}},n_{k,p_{\pm}}\propto\exp\left[-i\omega_{k,p_{\pm}}t\right]$,
where $\omega_{k}=\omega_{k,r}+i\gamma_{k}$ is the eigenvalue associated
with the unstable mode ($\gamma_{k}>0$) of the drift-wave instability,
we can write
\begin{equation}
\overline{n}_{q}=\left(A_{k,q}^{+}e^{-i\delta\omega t}+A_{k,q}^{-}e^{i\delta\omega t}\right)e^{(\gamma_{k}+\gamma_{p})t}\,,\label{eq:nqzonal}
\end{equation}
where $\delta\omega=\omega_{p,r}-\omega_{k,r}$ and $A_{k,p}^{\pm}$
are complex coefficients depending on the initial phase and amplitudes
of the modes $k$ and $p_{\pm}$. This suggests that the pertubation
is expected to grow at a rate $\gamma_{q}^{\delta n}=\gamma_{k}+\gamma_{p}$
initially.

Note that from figure \ref{fig:avalanche_growth} (b), we can estimate
the perturbation wavelength to be $\lambda_{q}=2(X_{+}-X_{-})\approx23.0$
(at $t\approx20$, when the distance between the bump and the hole
is still constant), which yields a wavenumber $q\approx0.27$. The
most unstable wavenumber can be calculated using the dispersion relation
(see Appendix \ref{sec:app-eigenvalues} and Ref. \citealp{gurcan:2022}
for the analytical expression and discussions) with $C=0.05$ and
$\kappa_{\ell}=5.0$, and yields $k_{y0}\approx0.39$. We can then
compute the perturbation growth rate $\gamma_{k}+\gamma_{p}\approx0.73$,
which corresponds to the dashed black line in figure \ref{fig:avalanche_growth}
(b), and which shows a very good agreement between this value and
the actual growth of $\delta n_{r}$ in the simulation.

The mechanism that determines the wavenumber $\lambda=2\pi/q$ of
the perturbation is yet to be found, but it might be similar to the
one that determines the typical size of the zonal flows, which remains
also unknown in the case of the Hasegawa-Wakatani system. Notice also
that the pair formation is only observed if the sharp gradient region
is sufficiently extended compared to $\lambda=2\pi/q$. If it is too
narrow, we do not see the formation of these large scale corrugations,
but rather smaller, less coherent structures.

\subsection{Relaxation and Spreading}\label{subsec:Spreading}

\subsubsection{Observations of profile relaxation and turbulent spreading in the
DNS }\label{subsec:spreadingDNS}

\begin{figure}
\centering{}\includegraphics[width=0.85\textwidth]{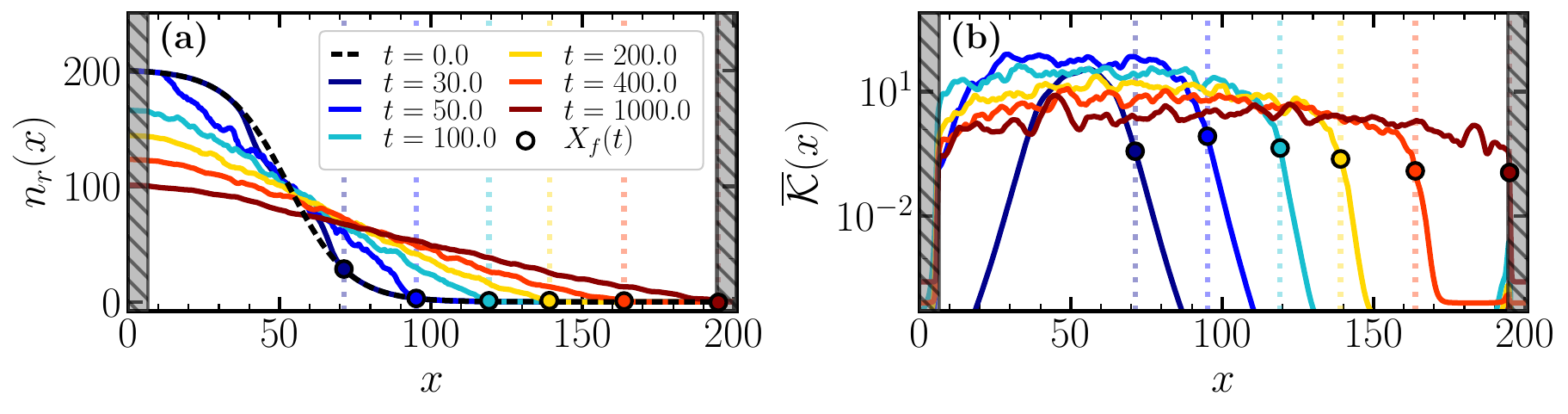}\caption{Left plot (a): radial density profile $n_{r}$ at different time steps
(the dashed line correspond to the initial profile). Right plot (b):
radial turbulent kinetic energy profile $\overline{\mathcal{K}}$
at the same time steps, in $y$ semilog plot. Circles correspond to
the rightmost position at which the radial kinetic energy $\overline{\mathcal{K}}$
is equal to $1\%$ of its maximum at a given time step (\ref{eq:Xfdef}),
which we use as a proxy for the turbulent front position $X_{f}(t)$.}\label{fig:spreading_times}
\end{figure}

In order to understand how the system continues to evolve, we consider
the spreading of the turbulent region together with the relaxation
of the radial density profile over longer time scales. In figure \ref{fig:spreading_times}
(a), we show the density profile $n_{r}$ at different time steps,
starting from $t=30$ (dark blue), when the initial pair consisting
of a bump and a hole collapses (the initial density profile is shown
in black dashed lines), until $t=1000$ (brown) when turbulence reaches
the right buffer and zonal flows start to accumulate and dominate
the system. A time animation of both density and zonal velocity radial
profiles is available on the repository page \citep[Movie 2]{guillon:2025b}.
It can be seen that the initial sharp gradient gradually relaxes in
time, as it extends to the right into the initially flat part of the
profile. Spreading to the right slows down, and stops when it reaches
the right buffer, at which point the system is completely covered
by turbulence, as can be seen in the right plots (d) and (h) in figure
\ref{fig:snapshots-spread}. Notice that the profile completely looses
its original shape and is very close to a straight line as a result
of the turbulent particle flux, flattening all the steep, large amplitude
initial perturbations.

We can also study the radial profile of the turbulent intensity and
its evolution in time. To do so, we define the radial turbulent kinetic
energy as:
\begin{equation}
\overline{\mathcal{K}}(x)\equiv\langle\widetilde{v}_{x}^{2}+\left(\widetilde{v}_{y}+\overline{v}_{y}\right){}^{2}\rangle_{y}\,,\label{eq:Kbardef}
\end{equation}
which is the poloidal average of the perturbed kinetic energy $e_{\mathcal{K}}(x,y)=\widetilde{v}_{x}^{2}+\left(\widetilde{v}_{y}+\overline{v}_{y}\right){}^{2}$,
where we include both zonal and non-zonal perturbations. The time
evolution of the radial profile of kinetic energy is shown in figure
\ref{fig:spreading_times} (b), at the same time steps with the density
profile on the left plot (a). The first time step $t=30$ (dark blue)
shows that the turbulent energy is peaked around $x=x_{a}$ where
the profile is the steepest. In fact, we could obtain this shape by
solving the equation
\begin{equation}
\partial_{t}\overline{\mathcal{K}}=2\gamma_{max}(\partial_{x}n_{r})\overline{\mathcal{K}}\,,\label{eq:growthradialkin}
\end{equation}
where $\gamma_{max}(\partial_{x}n_{r})$ is the maximum linear growth
rate computed by taking $\kappa(x)=-\partial_{x}n_{r}$ at each point
$x$, and is thus a function of the radial position. This is discussed
in more detail in section \ref{subsec:1Dspread} where we compare
the DNS results to those of a 1D reduced model of transport/turbulence
spreading. After the initial growth, the kinetic energy peak saturates
and spreads in the radial direction, together with the relaxing radial
density profile, decreasing in amplitude as the local free energy
source is lowered because the local profile gradient decreases. At
$t=1000$ (brown), turbulence completely covers the radial domain,
and we start to see some large-scale peaks at $x\approx40$ and $x\approx185$,
which corresponds to the formation of zonal flows. Notice that for
some intermediary time steps (\emph{e.g.} $t=100$, in lightblue),
we can observe traces of the numerical instability due to the mask
function gradient in the vicinity of the right buffer, as discussed
in section \ref{subsec:limitpen}. As mentioned, this instability
is 2-3 orders of magnitude lower than the turbulence level, and becomes
completely irrelevant once the turbulence reaches the right buffer.

In order to investigate how the density profile relaxation/turbulence
spreading front moves, we need a proxy for the position of the front
so that we can study its propagation. Ideally, the position of this
front should correspond to the point at which the density profile
is turbulent on the left, and still at rest on the right. To this
end, we use as a proxy the last radial point where the turbulent kinetic
energy exceeds a certain threshold, which we take as $1\%$ of the
maximum energy \emph{at each time step}:
\begin{equation}
X_{f}(t)\equiv\max_{x}\left\{ x\;:\;\overline{\mathcal{K}}(x,t)\geq0.01\times\max_{x}\left[\overline{\mathcal{K}}(x,t)\right]\right\} \,.\label{eq:Xfdef}
\end{equation}
 The position of $X_{f}$ at various time steps is shown in figure
\ref{fig:spreading_times} (coloured circles). We see that the proxy
provides a good estimate of the position of the front.\\

\begin{figure}
\centering{}\includegraphics[width=0.95\textwidth]{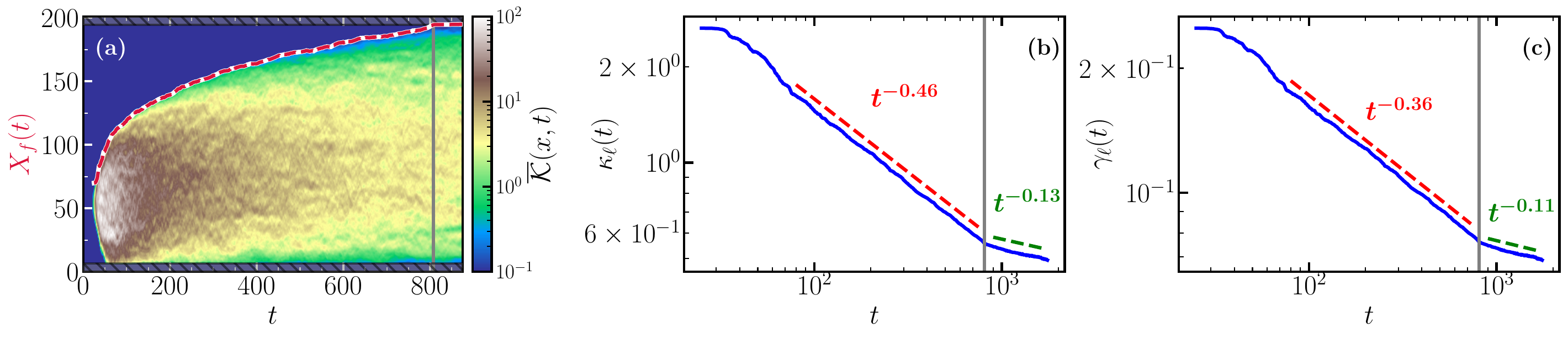}\caption{Left plot (a): time evolution of the turbulent front position $X_{f}(t)$
(dashed red line), on top of the spatiotemporal evolution of the radial
kinetic energy profile $\overline{\mathcal{K}}(x,t)$. Center plot
(b): time evolution of the left mean density gradient, defined by
Eq. \ref{eq:gammalt}. Right plot (c): time evolution of the corresponding
maximum growth rate, defined by Eq. \ref{eq:gammalt}. Both mean gradient
and growth rate are fit with power laws of the time (red and green
dashed lines). The vertical grey line denotes the time $t_{1}\approx800$
at which the front reaches the right buffer.}\label{fig:spreading_obs}
\end{figure}

In figure \ref{fig:spreading_obs} (a), we show the time evolution
of the turbulent front position $X_{f}(t)$ (dashed red line), on
top of the spatiotemporal evolution of the radial profile of kinetic
energy $\overline{\mathcal{K}}(x,t)$. We find again that the localised growth
of the turbulent energy around $x=x_{a}$ spreads quickly until $t=100$,
and then slows down but continues spreading at a slower rate, while
decreasing in amplitude. Note that we do not plot $X_{f}$ for $t<30$
because the definition of the proxy (\ref{eq:Xfdef}) fails when the
system is still in the linear instability phase, concentrated at $x_{a}$.
Around $t\approx600$, we start to see the formation of zonal flows,
evidenced by the yellow horizontal stripes in the spatiotemporal evolution
of $\overline{\mathcal{K}}$.

Using the turbulent front position, we can track the time evolution
of the mean density gradient to its left, $\kappa_{\ell}$ , which
corresponds to the slope of the high gradient part of the profile,
by using
\begin{equation}
\kappa_{\ell}(t)\equiv\frac{n_{r}(x_{b1},t)-n_{r}\left(X_{f}(t),t\right)}{X_{f}(t)-x_{b1}}\,,\label{eq:kappalt}
\end{equation}
which is defined using the difference between the left buffer position
$x_{b1}$ and the right turbulent front position $X_{f}(t)$. Note
that maybe we should rather define a left turbulent front position
and use it instead of $x_{b1}$. But since the left front reaches
the left buffer very quickly, it is practically equivalent to using
$x_{b1}$. The time evolution of $\kappa_{\ell}(t)$ is shown in figure
\ref{fig:spreading_obs} (b). After the initial collapse, the mean
left gradient decreases according to a time power law $t^{-0.46}$
(red dashed line) while turbulence spreads to the right. When the
turbulent front reaches the right buffer ($t\geq800$), $\kappa_{\ell}$
becomes the global density gradient of the system, and continues to
decrease according to a slower power law $t^{-0.13}$ (green dashed
line), due to the difference of particle flux between both ends of
the physical domain, as there is no particle source in these simulations
to compensate for the flux difference.

We can also compute the maximum growth rate in the left part associated
with $\kappa_{\ell}(t)$
\begin{equation}
\gamma_{\ell}(t)\equiv\gamma_{max}\left(\kappa_{\ell}(t)\right)\,,\label{eq:gammalt}
\end{equation}
which corresponds to the energy injection rate in the turbulent region.
The time evolution of $\gamma_{\ell}(t)$, shown in figure \ref{fig:spreading_obs}
(c), is very similar to that of $\kappa_{\ell}$, and also follows
two different time power laws before and after the front reaches the
buffer, though the power exponents slightly differ from that of $\kappa_{\ell}(t)$.
From the figure, we see that the energy injection rate decreases with
the relaxation of the profile. This can actually be used as an estimate
of the slow turbulent spreading velocity, which is actually rather
low for this system, as discussed in the next section.

\subsubsection{A 1D reduced model for coupled profile relaxation/spreading}\label{subsec:1Dspread}

In order to understand the key mechanisms that play a role in the
nature of profile relaxation/turbulence spreading that we observe
in the DNS without explicit source terms, we use a 1D reduced model
which features an equation on the radial turbulent kinetic energy
$\overline{\mathcal{K}}$ coupled with a transport equation for the
density profile $n_{r}$. The first equations is
\begin{equation}
\partial_{t}\overline{\mathcal{K}}=2\gamma_{max}(x,\partial_{x}n_{r})\overline{\mathcal{K}}-\beta_{NL}\overline{\mathcal{K}}^{2}+\chi_{\mathcal{\mathcal{K}}}\partial_{x}\left(\overline{\mathcal{K}}\,\partial_{x}\overline{\mathcal{K}}\right)\,,\label{eq:1DTKE}
\end{equation}
where in the RHS the first term corresponds to the local linear maximum
growth rate ($i.e.$ the growth rate of the most unstable mode $k_{y0}$),
computed using the radial density gradient $\partial_{x}n_{r}$ at
each point as the ``local'' background density gradient. The second
term corresponds to a nonlinear saturation due to eddy viscosity,
determined by $\beta_{NL}$, so that we would have a steady-state
$\overline{\mathcal{K}}(x)=2\gamma_{max}(x)/\beta_{NL}$ without turbulent
spreading. The latter is in turn represented by the last term on the
RHS, which is a nonlinear diffusion term, with a diffusion coefficient
$\chi_{\mathcal{K}}\overline{\mathcal{K}}$ that is proportional to
the turbulent kinetic energy itself. The second equation, for the
radial density profile, can be written as:
\begin{equation}
\partial_{t}n_{r}=D_{n}\partial_{x}\left(\overline{\mathcal{K}}\,\partial_{x}n_{r}\right)\,,\label{eq:1Dnr}
\end{equation}
where the turbulent flux is also assumed to have the same nonlinear
diffusivity as the turbulent kinetic energy, \emph{i.e.} $\Gamma_{n}=-D_{n}\overline{\mathcal{K}}\,\partial_{x}n_{r}$.
The coefficients $\beta_{NL},\chi_{\mathcal{K}}$ and $D_{n}$ are
the parameters of the model, which is very similar to previous reductions
introduced to study profile relaxation and turbulence spreading together.
Although here Eq. \ref{eq:1DTKE} is on the radial turbulent kinetic
energy $\overline{\mathcal{K}}$, for some of these works it is replaced
by the evolution of the turbulent intensity $\sum_{k}|\phi_{k}|^{2}$
denoted by $I$ \citep{hahm:2004,miki:2012}, $\varepsilon$ \citep{gurcan:2005}
or $E$ \citep{garbet:2007}. In Ref. \citealp{naulin:2005}, the
equation is on a general definition of the turbulent energy, written
$\epsilon$.

In order to compare the 1D reduced model to DNS, we incorporate the
boundary conditions of the penalisation method in Eqs. \ref{eq:1DTKE}
and \ref{eq:1Dnr} (see Appendix \ref{sec:app-1D}). Note that some
refined model \citep{miki:2012} can include an additional saturation
term due to zonal flow shearing, as well as an equation on the zonal
energy. However, as will be discussed in section \ref{subsec:Latest-times},
the fraction of zonal energy remains rather low during the spreading
phase, and the two equations system (\ref{eq:1DTKE}) and (\ref{eq:1Dnr})
seemed to be sufficient to reproduce the turbulent front spreading,
in our case. Additional refinements include a mean $E\times B$ poloidal
flow shearing effect, along with neoclassical diffusion and flow terms,
which are out of the scope of the present study.

\begin{figure}
\centering{}\includegraphics[width=0.85\textwidth]{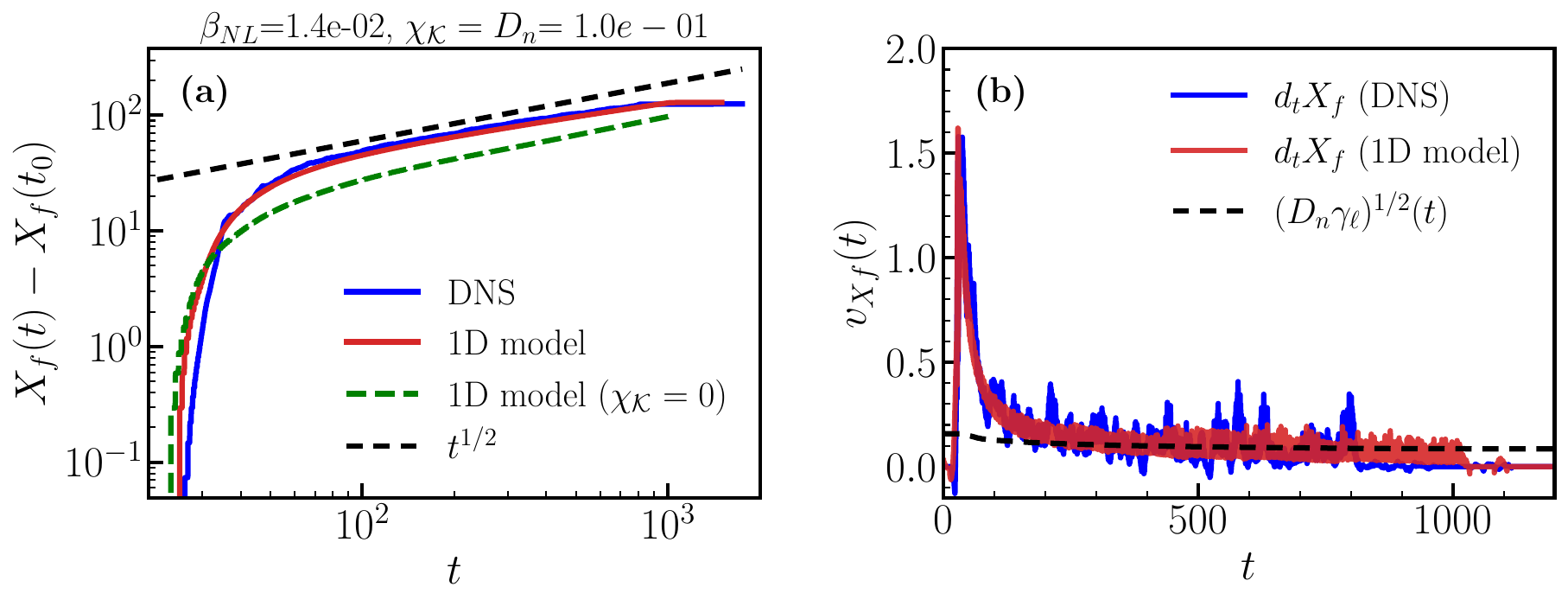}\caption{Comparison of the turbulent spreading between the DNS (blue) and the
1D reduced model (red). Left plot (a): log-log plot of the turbulent
front $X_{f}(t)$ computed using the proxy (\ref{eq:Xfdef}), from
which the initial front position $X_{f}(t_{0})$ has been subtracted.
The green dashed line corresponds to the 1D simulation without spreading
(\emph{i.e. }$\chi_{\mathcal{K}}=0$). The black dashed line correspond
to the slope of the diffusive spreading $X(t)\propto t^{1/2}.$ Right
plot (b): velocity of the turbulent front $v_{X_{f}}(t)=d_{t}X_{f}$
computed by differentiating $X_{f}(t)$ after applying a Savitzky-Golay
filter. The simulations are compared to the theoretical estimation
of the front velocity $\sqrt{D_{n}\gamma_{\ell}(t)}$ (dashed line)
where $\gamma_{\ell}$ is computed using Eq. \ref{eq:gammalt} and
$D_{n}=0.1$ is the diffusion coefficient of the 1D reduced density
transport equation (\ref{eq:1Dnr}).}\label{fig:Xf_DNS_vs_1D}
\end{figure}

In figure \ref{fig:Xf_DNS_vs_1D}, we show the comparison of the turbulent
front position $X_{f}(t)$ (a) and velocity $v_{X_{f}}(t)=d_{t}X_{f}$
(b) measured in both DNS and 1D reduced model. The numerical simulation
of the 1D reduced model is performed using $\beta_{NL}=0.014,\chi_{\mathcal{K}}=0.1$
and $D_{n}=0.1$ which yields satisfactory agreement with the DNS,
even though a formal optimisation study, which is left for future
work, could be used to improve this agreement. The turbulent front
position $X_{f}(t)$ is measured using the proxy (\ref{eq:Xfdef}),
from which we subtracted its initial value at $t_{0}=30$ (before
this time the proxy measurement fails because we are in the linear
phase). The velocity $v_{X_{f}}(t)=d_{t}X_{f}$ is measured in both
cases by differentiating $X_{f}$ after applying a Savitzky-Golay
filter \citep{savitzky:1964} with a window of 1000 time points and
a 3\textsuperscript{rd} degree polynomial, in order to filter out
the short fluctuations of the front position (which arise from the front position measurement using the proxy and from fast turbulent fluctuations), which correspond to abrupt and noisy
spikes in the velocity signal. That way, we can focus on the mean
velocity of the front which evolves on a slower time scale than that
of turbulence.

The comparison of the turbulent front behaviour between the DNS and
the 1D reduced model yields remarkably good agreement, especially
for the long time evolution of the front position $X_{f}(t)$, where
it seems to follow some kind of power law with $X_{f}(t)\propto t^{a}$,
where $a\approx0.4$. This suggests that turbulence spreading is slightly
subdiffusive in this particular example, since $X_{f}(t)$ evolves slower
than $t^{1/2}$, as shown by the black dashed line in the left plot
(a). Moreover, in different simulations of the 1D reduced model, we
oberve that the exponent $a$ of the power law scaling increases as
a function of $\chi_{\mathcal{K}}$ and $D_{n}$ (taken equal). 

To underline the importance of the turbulent spreading term $\chi_{\mathcal{\mathcal{K}}}\partial_{x}\left(\overline{\mathcal{K}}\,\partial_{x}\overline{\mathcal{K}}\right)$,
we performed a simulation of the 1D model taking $\chi_{\mathcal{K}}=0$,
while keeping the same values for $\beta_{NL}$ and $D_{n}$. The
front propagation of this simulation without the spreading term corresponds
to the green dashed line in figure \ref{fig:Xf_DNS_vs_1D} (a). We
see that in this case, the front does not propagate as far (2 times
less) as when turbulent spreading is included, and it really becomes
diffusive in this case (\emph{i.e.} evolves as \emph{$t^{1/2}$}).
To observe a similar front propagation, we need to multiply the diffusion
coefficient $D_{n}$ approximately by 10. In particular, we loose
the initial, brutal collapse of the profile ($t=50-100$ in figure
\ref{fig:spreading_obs} (a)). Hence, even though turbulence spreading
makes the front propagation slightly subdiffusive, its diffusion rate
becomes much (almost 10 times) higher than profile relaxation on its
own, suggesting that it plays a role in the front propagation in this
system.

The front velocity closely matches between the DNS and the reduced
model, although it fluctuates much more in the former due to the complete
turbulence description (with eddies intermittently surging into the
unperturbed region on the right), while it propagates more smoothly
in the 1D model where the spreading comes from the nonlinear diffusive
term. We compare the numerical results to the theoretical estimate
of the front velocity for a diffusive process $v_{front}=\sqrt{D_{n}\gamma_{\ell}(t)}$
\citep{gurcan:2005}, where $\gamma_{\ell}(t)$ is the maximum growth
rate in the turbulent domain (measured in the DNS), and $D_{n}$ is
the density coefficient from the reduced model. Although this estimation
fails in the early times, where the velocity decreases very rapidly,
it roughly matches both models on the long time, as evidenced by the
black line in figure \ref{fig:Xf_DNS_vs_1D} (b).

The animation of the energy and density profiles for both DNS and
1D model is available at the repository page \citep[Movie 3]{guillon:2025b},
and shows that the profiles can be different to some extent between
both simulations, even though the front evolution is very similar.
This could be improved by optimising the parameters of the 1D reduced
model using the DNS, in order to completely validate the comparison,
along with adding the effect of turbulence damping by zonal flows
shear, which is left for future work.

\subsection{Freezing of the system by zonal flows }\label{subsec:Latest-times}

\begin{figure}
\centering{}\includegraphics[width=0.95\textwidth]{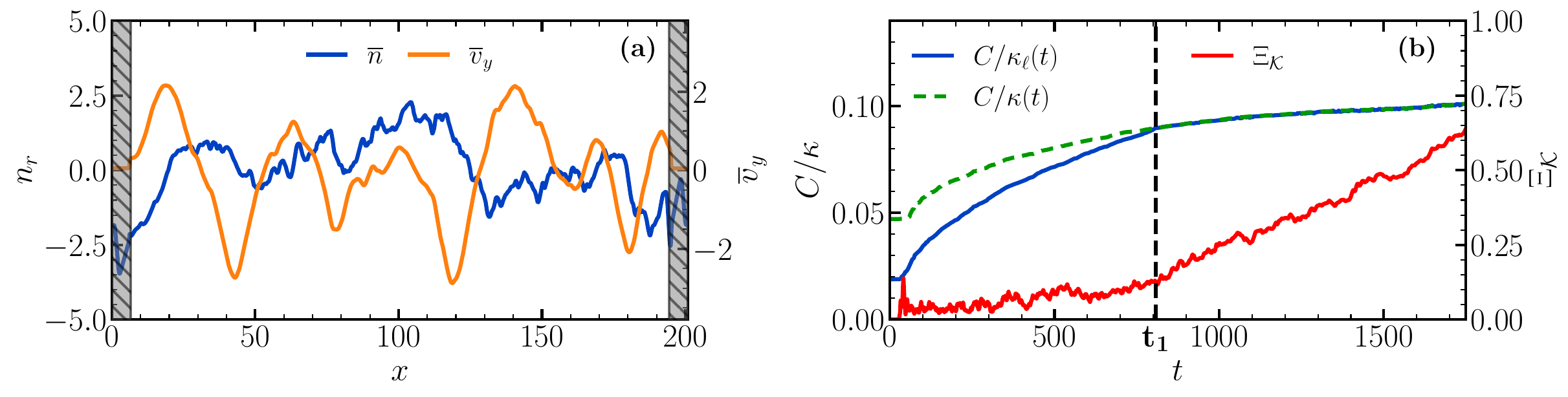}\caption{Left plot (a): zonal density profile $\overline{n}$ (blue) and zonal
velocity profile $\overline{v}_{y}$ (orange) at the last time step
of the simulation. Right plot (b): control parameter $C/\kappa$ computed
using the left density gradient $\kappa_{\ell}(t)$ (blue) and the
global mean gradient $\kappa$ (dashed green) as a function of time.
The fraction of zonal kinetic energy $\Xi_{\mathcal{K}}$ is also
shown in red. The vertical line marks the time $t_{1}\approx800$
at which the turbulent front reaches the right buffer. }\label{fig:spread_lsattimes}
\end{figure}

Last, we can discuss the final state of the system, after $t=t_{1}$,
at which the turbulent front reaches the right buffer. The density
profile continues to relax until it stops and develops large scale
corrugations. These corrugations can be seen as rising trends and sharp drops in the zonal density profile $\overline{n}$
(blue line), shown in figure \ref{fig:spread_lsattimes} (a), together with smaller scale fluctuations. At the same time, we see the
emergence of large scale zonal flows (orange) which become stationary.

This is due to the fact that, since the turbulent front has reached the right buffer, the density gradient $\kappa_{\ell}$,
which now corresponds to the mean global background $\kappa$ of the
system, is
close to the critical value $\kappa_{c}=0.5$ so that $C/\kappa_{c}=0.1$,
at which the fixed gradient system normally transitions to a zonal
flow dominated state \citep{numata:2007,grander:2024,guillon:2025}.
This is illustrated in figure \ref{fig:spread_lsattimes} (b) where
the ratios $C/\kappa_{\ell}(t)$ (blue) and $C/\kappa(t)$ (dashed
green) are shown. Because the zonal flows suppress the turbulent particle
flux, the radial density profile stops to evolve (actually it still
evolves a bit since the particle flux is not exactly zero, but extremely slowly). Indeed,
$C/\kappa$ becomes almost constant at the end of the simulation. Hence, we end up with a state that
is ``\emph{froze}n'' by the zonal flows. Note that, somewhat counter-intuitively,
because of the dependence of the adiabaticity parameter to temperature
through its dependence on parallel resistivity, the hotter the physical
system, the more easily it can be frozen by the zonal flows.

To quantify the level of zonal flows, we can measure the fraction
of zonal kinetic energy $\Xi_{\mathcal{K}}\equiv\mathcal{K}_{ZF}/\mathcal{K}$,
where $\mathcal{K}_{ZF}$ is the kinetic energy of zonal modes, and
$\mathcal{K}$ is the total kinetic energy of the system. The fraction,
shown in figure \ref{fig:spread_lsattimes} (b), starts from low values
at the beginning of the simulation, and reaches almost $70\%$ at
the end, even though it still continues to increase.

In some simulations with a sufficiently large 2D box, we observe that
zonal flow formation can be localised near the edges of the system, while
the center is still turbulent. Then, over time, zonal flows progressively
cover the domain and the turbulence becomes completely suppressed.
We leave the characterisation of this ``freezing front'', and its comparison with that of turbulence, as well
as the attempt to reproduce it using the 1D reduced model featuring
an equation on the zonal velocity/shear, to future studies. However, note that the phenomenon
is dependent on the value of $C$ and requires tailoring the initial
density profile to be observed.

\section{Particle source, sandpiles and stiffness}\label{sec:sources}

In this last section, we explore the behaviour of the system in the
presence of a particle source. In the radial density transport equation
(\ref{eq:hwfd-nr}), we add the following source term
\begin{equation}
S_{n}(\alpha)=\frac{\alpha}{\sigma_{n}\sqrt{2\pi}}\exp[-(x-x_{0})^{2}/(2\sigma_{n}^{2})]\,,\label{eq:Sndef}
\end{equation}
where $\alpha$ is the source amplitude (in this study $\alpha$ is
constant in time but it can be ramped), and $x_{0}$ and $\sigma_{n}$
are respectively its position and width. The source term has a Gaussian
shape, and is localised close to the left physical boundary so that
it mainly acts on the left part of the density profile. With this
source term, the particle budget inside the physical domain (\ref{eq:Nconserv})
becomes
\begin{equation}
d_{t}N_{\varphi}=\Gamma_{n}(x_{b1},t)-\Gamma_{n}(x_{b2},t)+P_{b2}(t)+\alpha\,,\label{eq:dtNphialpha}
\end{equation}
where we recall that $\Gamma_{n}(x_{b1},t)-\Gamma_{n}(x_{b2},t)$
is the difference between the particle fluxes at the two physical
boundaries, and $P_{b2}(t)=\int_{x_{b1}}^{x_{b2}}S_{b2}(x,t)\,dx$
is the integral over the physical domain of the sink $S_{b2}$ (\ref{eq:sourceartif2})
localised at $x_{b2}$ (because of the fixed boundary condition at
$x_{b2}$). Note that here
\begin{equation}
P_{b2}(t)\approx\sigma_{b}\sqrt{\frac{\pi}{2}}\partial_{x}\Gamma_{n}(x_{b2},t)\,,\label{eq:integratedsink2}
\end{equation}
assuming the artifical source is sufficiently narrow so that 
\[
\int_{x_{1}}^{x_{2}}\exp\left[-(x-x_{b2})^{2}/(2\sigma_{S_{b}}^{2})\right]dx\approx\sigma_{b}\sqrt{\frac{\pi}{2}}\,,
\]
and that $\partial_{t}n_{r}(x_{b2},t)\approx-\partial_{x}\Gamma_{n}(x_{b2},t)$
since the driving source $S_{n}(\alpha)$ is localised close to the
left edge.

We start the numerical simulation without the source term and with
a Gaussian initial profile (corresponding to $\kappa(t=0)\approx1.2$) with $C=0.05$,
with a padded resolution of $1024\times1024$. Details about the numerical
set-up and the source position and width are given in Appendix \ref{sec:app-source}.
After reaching the first saturation peak of the turbulent energy,
at time $t_{S}\approx207$, we stop the simulation and measure the
net outward particle flux, which gives $\Delta\Gamma\equiv\Gamma_{n}(x_{b2})-\Gamma_{n}(x_{b1})\approx1.2$
that we use as a reference value for setting the source amplitude.
Using the last time step of this saturated \emph{seed} simulation,
we start four different simulations which now include the source term
(\ref{eq:Sndef}), with the following four values of the source amplitude:
$\alpha=0$, $\alpha=\Delta\Gamma=1.2$, $\alpha=2\Delta\Gamma=2.4$
and $\alpha=10\Delta\Gamma=12$. An animation showing these four simulations
is available on the repository page \citep[Movie 4]{guillon:2025b}.

\begin{figure}
\centering{}\includegraphics[width=0.95\textwidth]{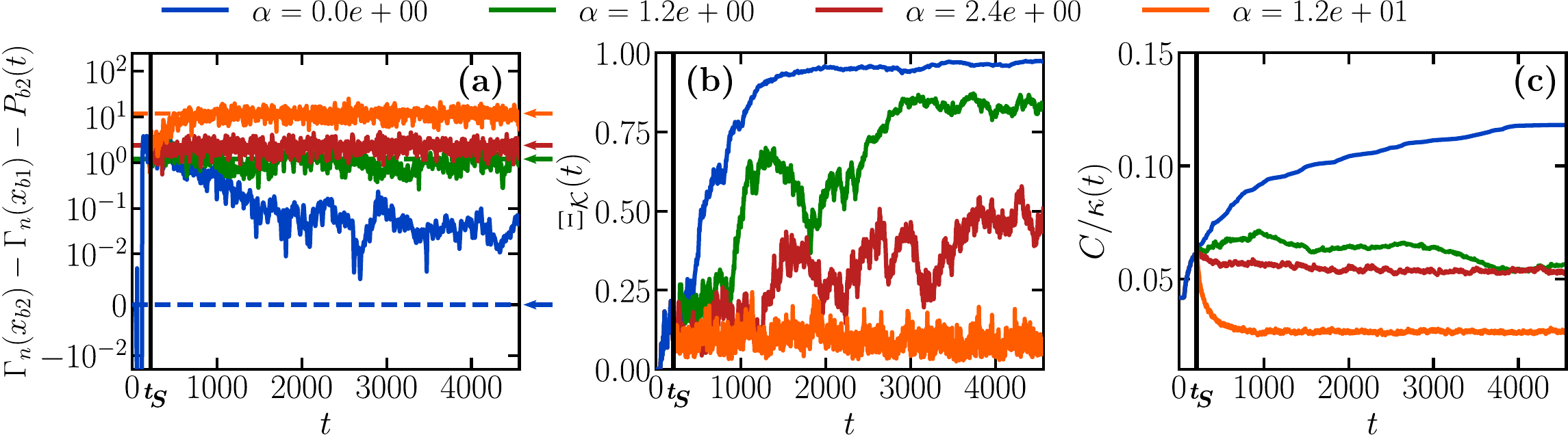}\caption{Comparison of observables between the four source amplitudes: $\alpha=0$
(blue), $\alpha=1.2$ (green), $\alpha=2.4$ (red) and $\alpha=12$
(orange). Left plot (a): outward particle flux $\Gamma_{n}(x_{b2},t)-\Gamma_{n}(x_{b1},t)-P_{b2}(t)$
in y symlog plot (filtered). Source amplitudes are indicated by dashed
lines and arrows on the right. Middle plot (b): zonal kinetic energy
fraction $\Xi_{\mathcal{K}}$. Right plot (c): control parameter $C/\kappa(t)$.
The vertical line at $t=t_{S}$ corresponds to the time at which the
sources are activated. }\label{fig:comp_sources}
\end{figure}

The time evolution of different observables are shown in figure (\ref{fig:comp_sources})
for the four source amplitudes ($\alpha=0$ in blue, $\alpha=1.2$
in green, $\alpha=2.4$ in red and $\alpha=12$ in orange). In the
left plot (a), the outward particle flux $\Gamma_{n}(x_{b2},t)-\Gamma_{n}(x_{b1},t)-P_{b2}(t)$
(accounting for the sink at $x_{b2}$) is shown, applying a Savitzky-Golay
filter with a window of 200 time points and a 3\textsuperscript{rd}
degree polynomial. The saturated value of the outward flux increases
with the source amplitude, and it relaxes towards zero in the absence
of source (blue line). Notice that the outward flux seems to fluctuate
around an equilibrium value close to the source amplitude $\alpha$,
as evidenced by the coloured dashed lines and arrows, except for the
decaying case (\emph{i.e. }$\alpha=0$ in blue), for which the flux
continues progressively to decrease. In this case, the profile will
collapse until there are either no particles or the flux is completely
suppressed by zonal flows. Note also that the actual non-filtered
flux (not shown here) fluctuates much more, and that it may become
momentarily negative, corresponding to narrow negative spikes in its
time evolution. This may be due to coherent structures getting reflected
off of the right buffer, due to the no-slip boundary condition acting
as a wall for coherent vortices.

Looking closely at the time evolution of both the fraction of zonal
kinetic energy $\Xi_{\mathcal{K}}(t)$ and the control parameter $C/\kappa(t)$,
which are shown respectively in figure (\ref{fig:comp_sources}) (b)
and (c), one finds that they display a very interesting phenomenon.
When there is no source (blue), the profile relaxes until $C/\kappa\sim0.1$
and the system becomes dominated by zonal flows, with $\Xi_{\mathcal{K}}\sim1$,
as discussed previously in section \ref{subsec:Latest-times}. At
the extreme opposite, for $\alpha=12$ (orange), $C/\kappa$ actually
decreases quite rapidly, since the density profile rises due to the
strong localised source, before it stabilises. Here the critical value
for the transition to zonal flows is never reached and the zonal fraction
remains very low.

For the two intermediate source levels, however, the final mean density
gradients are very close even though their turbulent states are very
different. For $\alpha=1.2$ (green), at first, the profile oscillates
between relaxing and rising, as evidenced by the upward and downward
evolution of $C/\kappa$, due to the source creating a stronger local
gradient, which generates more turbulence. Consequently, the profile
collapses due to the higher turbulent flux, and once it collapses,
the source makes it rise again. This cyclic behaviour is reminiscent
of the self-organised criticality (SOC) in sandpile models \citep{bak:1987,hwa:1992,carreras:1996},
where reaching a critical gradient makes the sandpile collapse until
it is again stable and starts to grow, oscillating around a marginal
equilibrium. What is interesting here, with the Hasegawa-Wakatani
system, is that the marginal stability in question is not a linear
stability, but a phase transition from a zonal flow dominated to a
turbulent state. Indeed, the critical gradient $\kappa_{c}$ such
that $C/\kappa_{c}=0.1$, at which the system transitions to zonal
flows and the particle flux is suppressed, defines a marginal state
towards which the system converges eventually (as discussed in section
\ref{subsec:Latest-times}), and around which the presence of a (weak)
particle source makes the profile oscillate. Indeed, we see that the
zonal fraction in figure (\ref{fig:comp_sources}) (b) for $\alpha=1.2$
goes above $50\%$ and then collapses when the profile rises again.
However at $t=2000$ the system is eventually able to definitely transition
to the zonal flow dominated state. Surprisingly, we see that the mean
gradient rises again and reaches that of the system with $\alpha=2.4$.
This can be explained by the hysteresis in the transition to zonal
flows \citep{guillon:2025}, in which zonal flows survive in the presence
of a larger mean gradient $\kappa>\kappa_{c}$ once they have formed.
Since zonal flows have reduced the particle flux, the source can
now efficiently feed the profile, which explains why $C/\kappa\approx0.05$
at the end of the simulation even though $\Xi_{\mathcal{K}}\approx80\%$.
The hysteresis allows to define a new marginal state for the system
where the mean gradient is higher. This is reminiscent of the effect
of mean shear flows in the L-H transition, which is explained by Ref.
\citealp{hinton:1993a} with a simple 1D model featuring hysteresis.
While self-organised criticality is studied for systems that exhibit
a 2\textsuperscript{nd} order phase transition, it is more relevant
here to speak about the so-called self-organised bistability \citep{gil:1996,disanto:2016}
which takes place at the onset of a 1\textsuperscript{st} order phase
transition with a hysteresis loop, and which can exhibit rare strong
events. The detailed study of this connexion is left for future work.

\begin{figure}
\centering{}\includegraphics[width=0.4\textwidth]{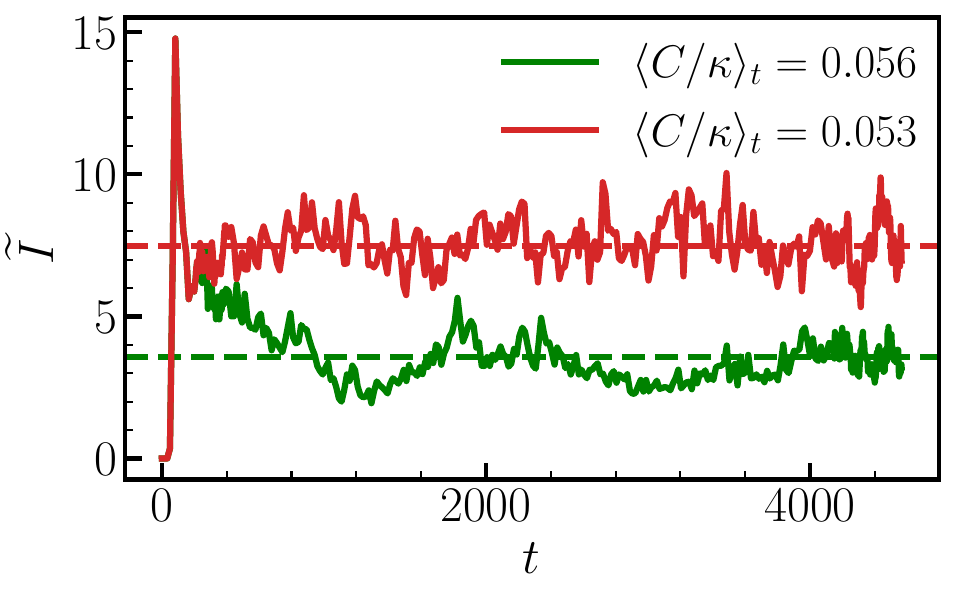}\caption{Time evolution of the non-zonal turbulent intensity $\widetilde{I}\equiv\sum_{k_{x},k_{y}\protect\neq0}|\phi_{k}|^{2}$
for $\alpha=1.2$ (green) and $\alpha=2.4$ (red). Mean values of
$\widetilde{I}$ (dashed lines) and of the control parameter $C/\kappa$
are taken over the last quarter of the simulations. }\label{fig:I_vs_t_Ckap}
\end{figure}

On the contrary, for $\alpha=2.4$ (red), the source is too strong
and does not enable the system to form zonal flows. Due to the strong
particle flux, the source cannot efficiently increase the density
gradient, which converges to that of $\alpha=1.2$. Hence both simulations
end up roughly with the same mean gradient, but completely different
zonal fractions, and completely different values of the fluxes. This
corresponds to profile stiffness observed in turbulent transport \citep{wolf:2003,garbet:2004,mantica:2009},
where small changes in the mean profile gradient results in dramatic
changes of the turbulent flux. This is further illustrated in figure
\ref{fig:I_vs_t_Ckap}, where we show the time evolution of the non-zonal
turbulent intensity $\widetilde{I}\equiv\sum_{k_{x},k_{y}\neq0}|\phi_{k}|^{2}$
for $\alpha=1.2$ (green) and $\alpha=2.4$ (red). Although there
is more than a factor two between the turbulence level in both simulations,
the value of the control parameter $C/\kappa$ (averaged on the last
quarter of the simulation) is almost the same. We argue that such
an observation is only possible in the parameter regime where the
hysteresis appears in the fixed gradient formulation.\\

\begin{figure}
\centering{}\includegraphics[width=1\textwidth]{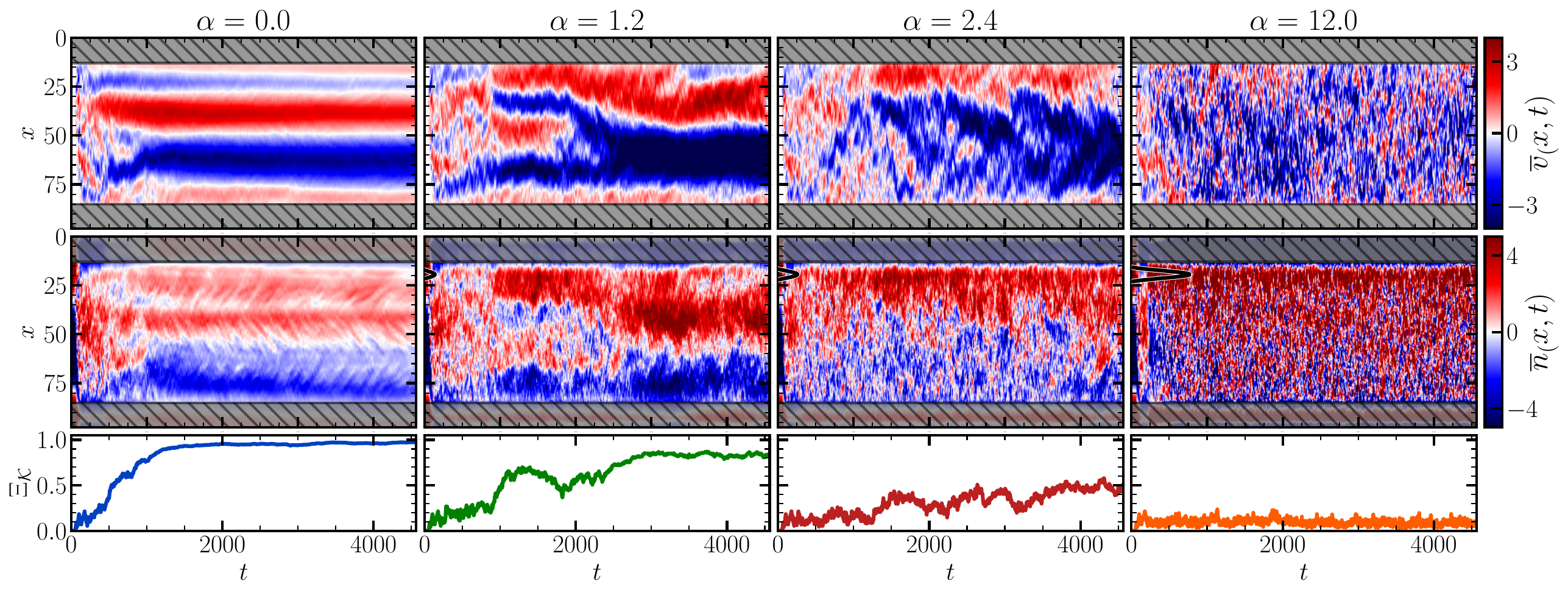}\caption{Hovmöller diagrams of the zonal velocity $\overline{v}_{y}$ (top
row) and zonal density $\overline{n}$ (middle row) profiles, for
the 4 source amplitudes $\alpha$ (the source is shown by the black
line on the density maps in arbitrary units). The bottom row shows
the time evolution of the zonal kinetic energy fraction $\Xi_{\mathcal{K}}$.}\label{fig:hovmoller}
\end{figure}

In order to further illustrate the differences between the dynamics of the system with these four different source amplitudes $\alpha$, we show
the Hovmöller diagrams of the zonal velocity $\overline{v}_{y}$ (top
row) and zonal density (middle row) $\overline{n}$ profiles, along with the corresponding
fraction of kinetic energy in zonal flows $\Xi_{\mathcal{K}}$
in the bottom row, in figure \ref{fig:hovmoller}. As expected, without any source ($\alpha=0$, the leftmost
column), which corresponds to the highest zonal energy fraction, we
see the formation of steady zonal flows, along with large scale density
corrugations. Conversely, for the strongest source amplitude ($\alpha=12$, the
rightmost column), both velocity and density profiles remain strongly
fluctuating, and we see no evidence of any long-lived zonal structure. Note
that it is possible to see the effect of the particle source (shown
in black line), localised at $x\approx20$, which corresponds to an
intense steady line (in time) on the spatiotemporal
evolution of the zonal density close to that position. For the intermediate source amplitude
($\alpha=2.4$, the 3\textsuperscript{rd} column from the left), the
velocity profile is less chaotic (especially in the end of the simulation)
compared to the previous case, which is consistent with the zonal
fraction being slightly below $50$\%. However, no clear pattern can be
observed, apart from the largest zonal mode available, and the zonal
density profile is still very chaotic. 

Finally, for the low source amplitude ($\alpha=1.2$, the 2\textsuperscript{nd}
column from the left), we can see the emergence of the zonal pattern
when the zonal fraction is higher than $50\%$, \emph{i.e} from $t\approx 1000$
to $t\approx 2000$ and after $t\approx 2500$. Between $t\approx 1800$ and $t\approx 2500$,
the fraction decreases below $50\%$ and then increases again, which
corresponds in figure \ref{fig:comp_sources} to the collapse of the
density profile, before it starts again to steepen with strong zonal
flows. Looking at the zonal velocity profile during this time interval,
it is not clear if the zonal structure actually collapses due to
the backtransition from zonal flows to turbulence, or if we only see
a merging of two flows (which can be accompanied by a short decrease
of the fraction that we observed in fixed-gradient simulations). It
seems however that this merging is chaotic, and we could really have
a breakup of the zonal structure. Furthermore, when zonal flows reform
after collapsing, they usually have a larger scale structure (see
also Ref. \citealp{grander:2024}), which is also the case here.

\section{Conclusion}

First results from the P-FLARE code, which performs numerical simulations
of flux-driven, reduced fluid models using the penalisation method,
were presented. The formulation relies on the premise that, by designing
buffer zones at the two extremes of the radial domain, where the fluctuations
are strongly damped, it is possible to modify the radial perturbations
so that they become periodic, allowing for the use of  fast Fourier
transforms. 

In order to demonstrate the capabilities of the code, it was applied
to the flux-driven Hasegawa-Wakatani model, where the relaxation of
a density profile that is locally steep in the left part of the radial
domain is considered. It was observed that, initially, as the drift-wave
instability saturates, it formed density bumps moving radially outward,
along with holes moving inward, before becoming unstable themselves.
Then, the relaxation of the profile on longer times, along with the
spreading of the turbulent region, were analysed. Using a 1D model
of two coupled equations on the radial kinetic energy and the density
profile, we obtained a qualitative and quantitative agreement with
the DNS for the turbulent front propagation, and found that the spreading
was slightly subdiffusive in this specific example.
In the present study, the parameters for the 2D model were tuned by
hand to match the DNS, which can be extended in the future using
an automated optimisation of the parameters based on the DNS data
\citep{kobayashi:2015}. Currently, the code has zonal flows, but
no mean flow evolution. We also intend to implement mean flow profile
evolution, and compare with 1D models that have zonal and mean flows
and can describe L-H transition \citep{miki:2012,chone:2015}. 

In this study, we also observed that the critical mean gradient $\kappa_{c}$
such that $C/\kappa_{c}\approx0.1$ defines some kind of marginally stable
state, towards which the system converges if we have initially $C/\kappa<0.1$.
We have argued that this was a consequence of the transition between a 2D
turbulent state to a zonal flow dominated state that is observed in
the fixed gradient system with varying $C/\kappa$. When zonal flows
form, they suppress the turbulent transport and thus the profile relaxation slows down, to a halt. This is possible since neither large nor small-scale dissipations have been applied on zonal flows in this model. 

The impact of the transition was further explored by adding a localised
particle source. It was observed that, close to the transition point,
the system exhibits behaviour similar to sandpile models studied
in self-organised criticality. However in our case the stable state
was dominated by zonal flows, and became turbulent only when the particle
source pushed the mean gradient beyond the transition point. If the
source amplitude was sufficiently low, the profile would oscillate
close to the critical slope $\kappa_{c}$. Notice that, since the transition
in the fixed gradient version of the Hasegawa-Wakatani system shows
characteristic features of a 1\textsuperscript{st} order phase transition,
such as the existence of a hysteresis loop, we might actually have
self-organised bistability, which needs to be further characterised.

One consequence of this hysteresis is the observation of profile stiffness.
For very similar mean density gradients, we observed  that both the
turbulent particle flux and the turbulent intensity were different
by a factor of two, corresponding to very different zonal flow levels.
It seems that, with a small amplitude source, the system is able to
transition to zonal flows, which suppress the transport. The particle
source can then efficiently steepen the profile without zonal flows
immediatly collapsing, due to the bistability implied by the hysteresis
loop. This mechanism needs to be studied in more detail and extended
to other fluid systems which may exhibit a similar transition, \emph{e.g.}
ITG turbulence close to the Dimits shift \citep{dimits:2000}.

While this first study using P-FLARE was confined to the flux-driven
Hasegawa-Wakatani system, the code can be easily adapted to other
reduced fluid models \citep{horton:1981,horton:1988,sarazin:1998,ivanov:2020,panico:2025},
or using the gyro-moment approach \citep{frei:2023,hoffmann:2023}.
Future perspectives include implementing self-consistent evolution
of a mean $E\times B$ velocity profile and its damping towards a
neoclassical profile, which could be used to study other phenomena
such as the L-H transition. Benchmarking flux-driven reduced models,
and performing detailed comparisons between flux-driven and fixed
gradient simulations especially near the threshold of zonal flow formation, are also considered.

\section*{Acknowledgments}
This work was granted access to the Jean Zay super-computer of IDRIS
under the allocation AD010514291R2 by GENCI. This work has been carried
out within the framework of the EUROfusion Consortium, funded by the
European Union via the Euratom Research and Training Programme (Grant
Agreement No 101052200 --- EUROfusion) and within the framework of
the French Research Federation for Fusion Studies.

\bibliographystyle{unsrturl}
\bibliography{plguillon}

\appendix

\section{Benchmark of the relaxation/spreading simulation with BOUT++}\label{sec:Benchmark}

In order to compare both runtimes and physical results with more conventionnal
finite difference scheme, we also performed a benchmark of the relaxation
simulation with the BOUT++ code \citep{dudson:2009}. To make the
comparison possible, the domain simulated in BOUT++ is set to correspond
to the physical domain available in P-FLARE, \emph{i.e.} keeping the
full poloidal domain and removing the buffer zone. Hence, to compare
with a $1024\times1024$ padded P-FLARE simulation with the buffer
settings given in table \ref{tab:bench_params}, we use a resolution
$N_{x}\times N_{z}=506\times682$ with BOUT++ (the $z$ axis corresponds
to $-y$ in P-FLARE). Hence the $x$ axis in BOUT++ should be understood
as being shifted by the left buffer position $x_{b1}$ from P-FLARE.
All parameters for both P-FLARE and BOUT++ simulations are given in
table \ref{tab:bench_params}. We fix $\kappa=0$
and use, instead, the initial density profile in BOUT++. Note that we use the same (modified) Hasegawa-Wakatani equations as
P-FLARE, \emph{i.e. }with classical (and not hyper) viscosity, acting
only on non-zonal fluctuations, in BOUT++.

\begin{table}
\begin{centering}
\begin{tabular*}{0.99\textwidth}{@{\extracolsep{\fill}}>{\centering}m{0.20\textwidth}>{\centering}m{0.16\textwidth}>{\centering}m{0.05\textwidth}>{\centering}m{0.12\textwidth}>{\centering}m{0.05\textwidth}>{\centering}m{0.05\textwidth}>{\centering}m{0.16\textwidth}}
\toprule
\multicolumn{4}{c}{Physical parameters} & \multicolumn{3}{c}{Initial profile}\vspace{0.2cm}\tabularnewline
$L_{x},L_{z}$ (BOUT++) & $L_{x},L_{y}$ \hspace{0.2cm}(P-FLARE) & $C$ & $\nu=D$ & $\kappa_{\ell}$ & $\alpha$ & $x_{a}$\vspace{0.2cm}\tabularnewline

$32\pi-2x_{b1}$, $32\pi$ & $32\pi$, $32\pi$ & $0.05$ & $6.7\times10^{-2}$ & $10$ & $2$ & $1.8x_{b1}=23.90$\tabularnewline
\addlinespace[0.2cm]
\end{tabular*}
\par\end{centering}
\centering{}%
\begin{tabular*}{0.99\textwidth}{@{\extracolsep{\fill}}>{\centering}p{0.06\textwidth}>{\centering}p{0.06\textwidth}>{\centering}p{0.06\textwidth}>{\centering}p{0.06\textwidth}>{\centering}p{0.06\textwidth}>{\centering}p{0.06\textwidth}>{\centering}p{0.2\textwidth}>{\centering}p{0.2\textwidth}}
\midrule
\multicolumn{8}{c}{Penalisation parameters}\vspace{0.2cm}\tabularnewline
 
\multicolumn{3}{c}{Buffer zone} & \multicolumn{3}{c}{Smoothing function} & Penalisation coefficient & Artificial source size\vspace{0.2cm}\tabularnewline
$x_{b1}$ & $x_{b2}$ & $\delta x_{b}$ & $x_{m1}$ & $x_{m2}$ & $\delta x_{m}$ & $\mu$ & $\sigma_{S_{b}}$\vspace{0.2cm}\tabularnewline

13.27 & 87.26 & 8.84 & 6.63 & 93.90 & 5.90 & 100 & $5L_{x}/N_{x}=0.74$\tabularnewline
\bottomrule
\end{tabular*}\caption{Top: parameters for BOUT++ and P-FLARE simulations. Bottom: penalisation
parameters for P-FLARE.}\label{tab:bench_params}
\end{table}

As for initial conditions, the density profile is initialised with
the hyperbolic tangent formula (Eq. \ref{eq:nr0_spread}), with $\kappa_{\ell}=10$,
$x_{a}=1.8x_{b1}$ and $\alpha=2$ so that the steep region is shorter
in order to observe enough spreading to the right (in BOUT++ the profile
is shifted by $x_{b1}$ to the left). All non-zonal density fluctuations
are set to 0. The initial vorticity fluctuations are set using the
\texttt{mixmode} function from BOUT++ with an amplitude of $10^{-3}$.
In practice, we take the initial conditions from BOUT++ and use them
in P-FLARE. The vorticity fluctuations are then extrapolated to their
mean value inside the buffer zone, and the density profile is extended
in that region (using the analytical formula Eq. \ref{eq:nr0_spread}).
Both simulation are run until $t=200$.

The boundary conditions in BOUT++ are Neumann at the inner radial
axis, Dirichlet at the outer radial axis, and periodic along the poloidal
axis. Note that, as mentioned in section \ref{subsec:BC-radial},
we introduced in P-FLARE an artificial sink localised at $x_{b2}$,
in order to have the equivalent of a Dirichlet boundary condition
at the outer radial axis. Since the source has a finite width, it
allows for the particle flux to leave the radial domain. On the contrary
in BOUT++, the Dirichlet boundary condition is applied only at the
very edge. Consequently, we observed in some BOUT++ simulations that
once turbulence has reached the right edge, the particles pile up
at the right end of the domain, and the density profile jumps from
0 at the very edge (where the Dirichlet BC is applied), to a higher
value inside the radial domain. Hence, a sink with finite width similar
to that used in P-FLARE should probably also be implemented in BOUT++, in order
to allow particles exiting the domain, but this is beyond the scope
of this work. 

\begin{figure}
\centering{}\includegraphics[width=0.8\textwidth]{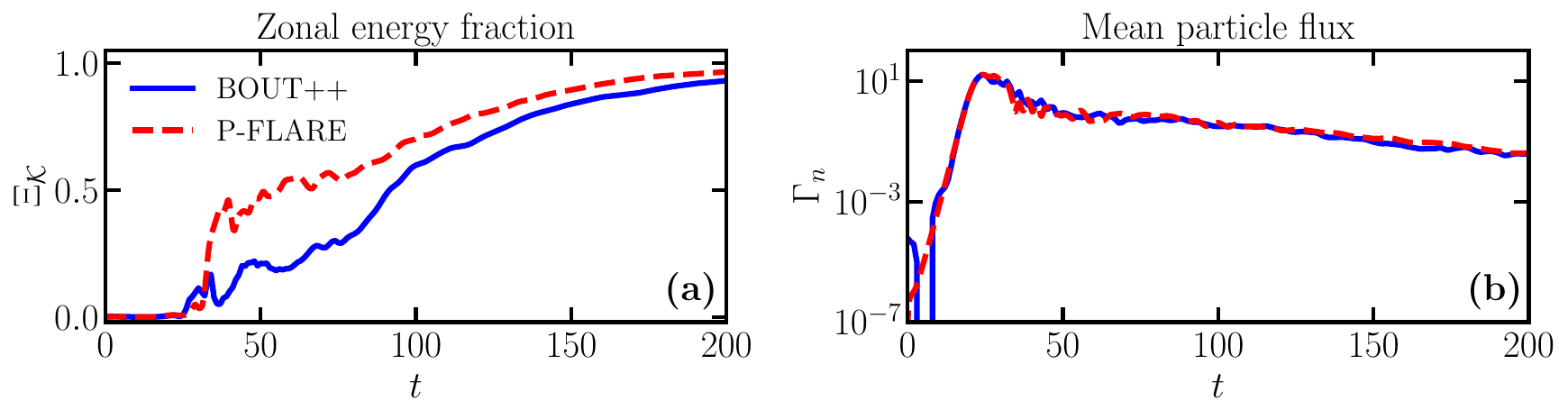}\caption{Zonal kinetic energy fraction $\Xi_{\mathcal{K}}$ (a) and mean radial
particle flux $\Gamma_{n}$ (b) versus time obtained with BOUT++ (plain
blue) and P-FLARE (dashed red) simulations.}\label{fig:BOUT_0D}
\end{figure}

As a first result, we show in figure \ref{fig:BOUT_0D} the time evolution
of the zonal kinetic energy fraction (left plot) and the mean radial
particle flux (right plot) for both codes. The time evolutions of the
fluxes are almost identical for both codes. The zonal fractions differ
between $t=50$ and $t=100$, with zonal flows forming earlier in
the P-FLARE simulation, but otherwise they display reasonably close values.
The difference in the transient evolution of zonal fractions can be
explained by the fact that finite difference introduces numerical
dissipation due to the error made when computing derivatives, and
this dissipation is applied to both zonal and non-zonal fluctuations,
which might affect the formation of zonal flows. On the contrary,
no such numerical dissipation is present when using FFTs, which compute
derivatives almost exactly. Nevertheless, both codes predict the same
final level of zonal flows and particle flux, which is eventually
the goal of such codes.

\begin{figure}
\centering{}\includegraphics[width=1\textwidth]{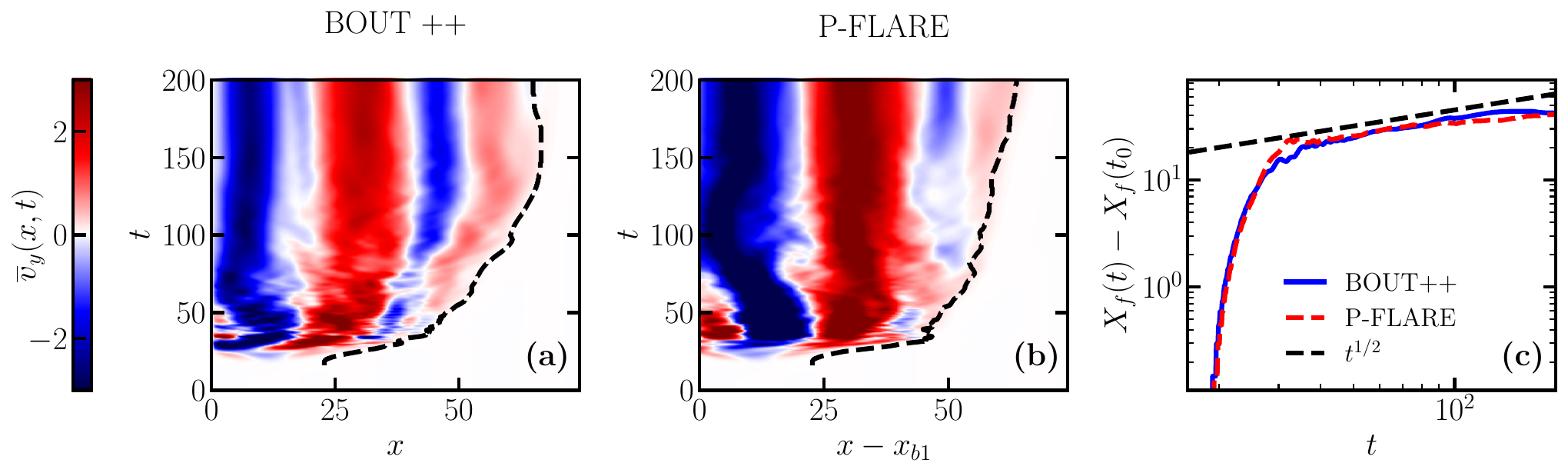}\caption{Hovmöller diagrams of the zonal velocity $\overline{v}_{y}$ profiles
obtained using BOUT++ (a) and P-FLARE (b), with the turbulent front
proxy $X_{f}(t)$ in dashed black. The time evolution of the turbulent
front position is shown in the right plot (c) obtained with BOUT++
(plain blue) and P-FLARE (dashed red), in log-log. Note that for the
P-FLARE results, we shifted the $x$ axis and the front position $X_{f}$
by the left buffer position $x_{b1}$, to line up with BOUT++.}\label{fig:BOUT_front}
\end{figure}

In figure \ref{fig:BOUT_front}, we compare the spatiotemporal evolutions
of the zonal velocity profile obtained with BOUT++ (a) and P-FLARE
(b). The profiles are very similar, spread towards the outer radial
axis at similar rates, and display the formation of similar zonal
structures. We can then look at the turbulent front propagation, which
we compute using the same proxy as in Eq. \ref{eq:Xfdef}. In the
right plot (c), we show the time evolution of the fronts $X_{f}$
(from which the initial front position $X_{f}(t_{0})$ has been subtracted)
obtained using BOUT++ (plain blue) and P-FLARE (dashed red). Both
fronts exhibit very similar propagation, even though the front from
BOUT++ seems to spread slightly faster. This can again be explained
by the additional numerical dissipation from finite difference, which
may increase the diffusion rate and introduce numerical damping on zonal flows.

Therefore, BOUT++ and P-FLARE exhibit very similar results for the
case of the relaxation study, which underlines the consistency of
the boundary condition implementation in P-FLARE, compared to using
Neumann at the left boundary and Dirichlet at the right in a finite
difference scheme. 

Comparing runtimes is not straightforward, since BOUT++ was parallelised
on 2 CPUs, while P-FLARE was run on a V100 GPU. In this case, it took
30 min for BOUT++ versus 40 min for P-FLARE. However, this actually depends on the performance of the solver and the numerical tolerances used in the adaptive steep computation. Moreover, we expect to
gain computational speed using FFTs when going to higher resolution
(\emph{i.e.} $4096^{2}$) compared to finite difference, and with
using a solver which accounts for stiffness (which is the case for
diffusion in BOUT++). Furthermore, what we advocate with this new
method is also the accuracy provided by FFTs: we can afford resolving
a wide range of scales without the effect of numerical dissipation.
Finally, the method allows for more flexibility, especially
regarding developing reduced models based on Fourier space properties
of turbulence (\emph{e.g. }Large Eddy Simulation, triad truncations,
etc.) which is eventually one of our goals with developing such a
code. 

\section{Time evolution of energy, flux and density for the relaxation/spreading
simulation}\label{sec:0Drelax}

\begin{figure}
\centering{}\includegraphics[width=0.85\textwidth]{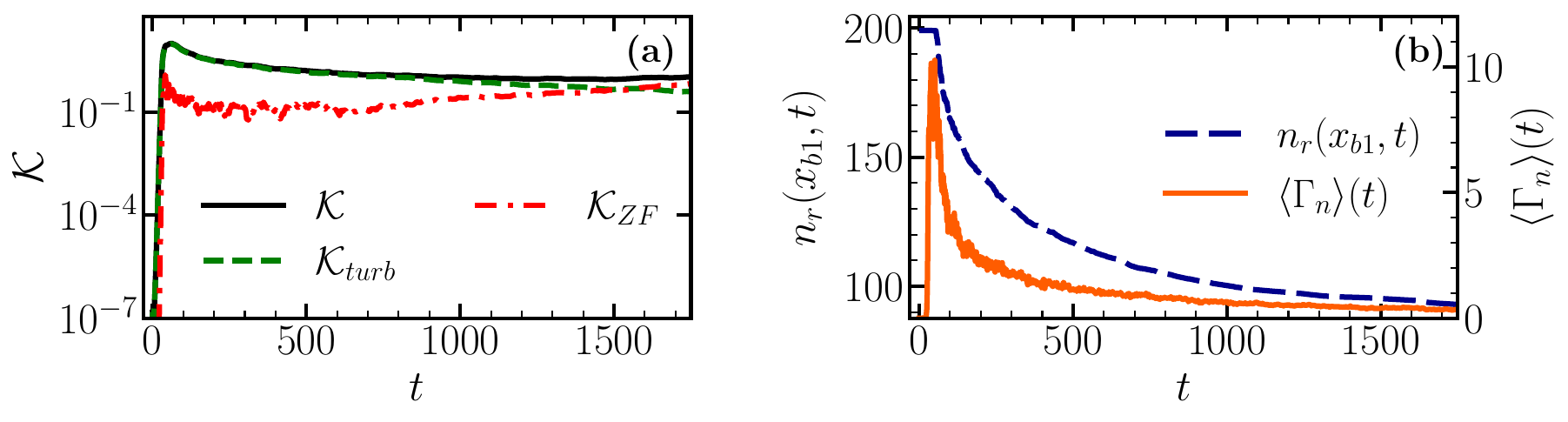}\caption{Time evolution of some quantities for the relaxation simulation (see
section \ref{sec:spread_transport}). Left plot (a): time evolution
of the total (black), zonal (red dashdot) and non-zonal (green dashed)
kinetic energy averaged over the physical domain. Right plot (b):
time evolution of the density at the left boundary $n_{r}(x_{b1},t)$
(dashed blue) and of the radial particle flux $\langle\Gamma_{n}\rangle$
(orange) averaged over the physical domain.}\label{fig:0Drelax}
\end{figure}

In figure \ref{fig:0Drelax}, we show the time evolution of several
quantities for the $4096\times4096$ padded resolution simulation
used for the study of profile relaxation and turbulence spreading
(see section \ref{sec:spread_transport}). In the left plot (a), we
show the total kinetic energy $\mathcal{K}$ (black), along with its
zonal component $\mathcal{K}_{ZF}$ (dashdot red) and its non-zonal
component $\mathcal{K}_{turb}$, averaged over the physical domain
$[x_{b1},x_{b2}]\times L_{y}$. In the right plot (b), we show the
time evolution of the density at the left boundary $n_{r}(x_{b1},t)$
(dashed blue), along with the radial particle flux $\langle\Gamma_{n}\rangle$
(orange) averaged over the physical domain.

\section{Eigenvalues of the Hasegawa-Wakatani system }\label{sec:app-eigenvalues}

Linearisation and Fourier transform of the Hasegawa-Wakatani equations
(\ref{eq:hw-om}) and (\ref{eq:hw-n}) for non-zonal modes $(k_{y}\neq0)$
yields \citep{gurcan:2022}
\begin{subequations}
\begin{align}
\partial_{t}\phi_{k}+(A_{k}-B_{k})\phi_{k} & =\frac{C}{k^{2}}n_{k}\,,\label{eq:HWappphik}\\
\partial_{t}n_{k}+(A_{k}+B_{k})n_{k} & =(C-i\kappa k_{y})\phi_{k}\text{ ,}\label{eq:HWappnk}
\end{align}
\end{subequations}
 where
\begin{subequations}
\begin{align}
A_{k}= & \frac{1}{2}\left[(Dk^{2}+C)+\left(\frac{C}{k^{2}}+\nu k^{2}\right)\right]\,,\label{eq:HWappAk}\\
B_{k}= & \frac{1}{2}\left[(Dk^{2}+C)-\left(\frac{C}{k^{2}}+\nu k^{2}\right)\right]\,.\label{eq:HWappBk}
\end{align}
\end{subequations}

The two eigenvalues $\omega_{k}^{\pm}(C,\kappa,\nu,D)=\omega_{k,r}^{\pm}+i\gamma_{k}^{\pm}$
can then be written as
\begin{equation}
\omega_{k}^{\pm}=\Omega_{k}^{\pm}-iA_{k}\,\label{eq:omkapp}
\end{equation}
 with
\begin{equation}
\Omega_{k}^{\pm}=\pm\left(\sigma_{k}\sqrt{\frac{H_{k}-G_{k}}{2}}+i\sqrt{\frac{H_{k}+G_{k}}{2}}\right)\,,\label{eq:Omkpmapp}
\end{equation}
 where $\sigma_{k}=\text{sign}(\kappa k_{y})$,
\begin{equation}
H_{k}=\sqrt{G_{k}^{2}+C^{2}\kappa{{}^2}k_{y}^{2}/k^{4}}\,,\label{eq:Hkapp}
\end{equation}
and
\begin{equation}
G_{k}=\left(B_{k}^{2}+\frac{C^{2}}{k^{2}}\right)\,.\label{eq:Gkapp}
\end{equation}

\section{Complete 1D reduced model for turbulent spreading}\label{sec:app-1D}

We detail the complete 1D reduced model used for reproducing the profile
relaxation and the turbulent spreading in section \ref{subsec:1Dspread}.
The complete equations are
\begin{subequations}
\begin{align}
\partial_{t}\overline{\mathcal{K}} & =2\gamma_{max}(x,\partial_{x}n_{r})\overline{\mathcal{K}}-\beta_{NL}\overline{\mathcal{K}}^{2}+\chi_{\mathcal{\mathcal{K}}}\partial_{x}\left(\overline{\mathcal{K}}\,\partial_{x}\overline{\mathcal{K}}\right)-\mu H(x)\overline{\mathcal{K}}\,,\label{eq:1DappKbar}\\
\partial_{t}n_{r} & =D_{n}\partial_{x}\left(\overline{\mathcal{K}}\,\partial_{x}n_{r}\right)+S_{b2}(x,t)-\mu H(x)\left[n_{r}(x,t)-n_{buff}(x,t)\right]\,,\label{eq:1Dappnr}
\end{align}
\end{subequations}
where we introduced the same penalisation terms $\mu H(x)$ as in
Eqs. \ref{eq:hw-ubar-pen} and \ref{eq:hw-nr-pen}, along with the
artificial source (\ref{eq:sourceartif2}) in the density equation.
The parameters are the same as for the DNS (see table \ref{tab:spread_params}),
except that we use a resolution of $N_{x}=682$ in this case.

Numerical simulations of the 1D reduced model are performed using
a finite-difference scheme and integrated in time using the \texttt{RK45}
method from \texttt{scipy.integrate.solve\_ivp} with an adaptative
time step. We use the same initial conditions as for the DNS, as well
as the same time steps for saving the solution.

\section{Numerical set-up with particle sources}\label{sec:app-source}

We perform $1024\times1024$ padded resolution simulations. We start
the \emph{seed} simulation with the following initial profile:
\begin{equation}
n_{r}(x,t=0)=L_{x}\exp\left[-4(x/L_{x})^{2}\right]\,,\label{eq:sourceappnr0}
\end{equation}
and we impose the following particle source:
\begin{equation}
S_{n}(\alpha)=\frac{\alpha}{\sigma_{n}\sqrt{2\pi}}\exp[-(x-x_{0})^{2}/(2\sigma_{n}^{2})]\,.\label{eq:sourceappsource}
\end{equation}
All physical, sources and penalisation parameters are given in table
\ref{tab:paramssource}.

\begin{table}
\begin{centering}
\begin{tabular*}{0.99\textwidth}{@{\extracolsep{\fill}}>{\centering}m{0.12\textwidth}>{\centering}m{0.05\textwidth}>{\centering}m{0.12\textwidth}>{\centering}m{0.18\textwidth}>{\centering}m{0.18\textwidth}}
\toprule
\multicolumn{3}{c}{Physical parameters} & \multicolumn{2}{c}{Source parameters}\vspace{0.2cm}\tabularnewline
$L_{x},L_{y}$ & $C$ & $\nu=D$ & $x_{0}$ & $\sigma_{n}$\tabularnewline

$97.8$, $97.8$ & $0.05$ & $8.8\times10^{-3}$ & $1.5x_{b1}=19.35$ & $L_{x}/50=1.95$\vspace{0.2cm}\tabularnewline
\midrule 
\end{tabular*}
\begin{tabular*}{0.99\textwidth}{@{\extracolsep{\fill}}>{\centering}p{0.06\textwidth}>{\centering}p{0.06\textwidth}>{\centering}p{0.06\textwidth}>{\centering}p{0.06\textwidth}>{\centering}p{0.06\textwidth}>{\centering}p{0.06\textwidth}>{\centering}p{0.2\textwidth}>{\centering}p{0.2\textwidth}}
\multicolumn{8}{c}{Penalisation parameters}\vspace{0.2cm}\tabularnewline
\multicolumn{3}{c}{Buffer zone} & \multicolumn{3}{c}{Smoothing function} & Penalisation coefficient & Artificial source size\vspace{0.2cm}\tabularnewline
$x_{b1}$ & $x_{b2}$ & $\delta x_{b}$ & $x_{m1}$ & $x_{m2}$ & $\delta x_{m}$ & $\mu$ & $\sigma_{S_{b}}$\vspace{0.2cm}\tabularnewline

12.90 & 84.87 & 8.60 & 6.45 & 91.32 & 5.73 & 100 & $5L_{x}/N_{x}=0.72$\tabularnewline
\bottomrule
\end{tabular*}
\par\end{centering}
\caption{Physics and penalisation parameters for the $1024\times1024$ padded
simulations with a source term.}\label{tab:paramssource}
\end{table}

\end{document}